\documentclass[12pt]{article}
\usepackage[utf8]{inputenc}
\usepackage[DIV=13]{typearea}\usepackage{ragged2e}
\usepackage{calligra,amsmath,amsfonts,bbm,mathrsfs,amssymb}
\usepackage{slashed,cancel}
\usepackage{cite}
\usepackage{bm} 
\usepackage{hyperref}
\hypersetup{colorlinks=true,urlcolor=magenta,anchorcolor=blue,citecolor=blue,filecolor=blue,linkcolor=magenta,menucolor=blue,linktocpage=true,pdfproducer=medialab}
\usepackage[english]{babel}
\usepackage{catchfile}
\usepackage{indentfirst}
\usepackage{graphicx}
\usepackage{float}
\usepackage{enumerate}
\usepackage{caption}
\usepackage{subcaption}
\usepackage{multirow}
\usepackage{tikz-feynman}
\usepackage[normalem]{ulem} 
\usepackage{booktabs}
\usepackage{physics}
\usepackage{url}
\usepackage{soul}
\newcommand{\email}[1]{\href{mailto:#1}{\tt #1}}

\numberwithin{equation}{section}

\newcommand{\blue}[1]{\color{blue} #1 \color{black}}
\newcommand{\magenta}[1]{\color{magenta} #1 \color{black}}
\newcommand{\be}{\begin{equation}}
\newcommand{\ee}{\end{equation}}
\newcommand{\ba} {\begin{equation}\begin{aligned}}
\newcommand{\ea} {\end{aligned}\end{equation}}
\newcommand{\bea}{\begin{eqnarray}}
\newcommand{\eea}{\end{eqnarray}}

\newcommand{\sL}{\mathscr{L}}

\newcommand{\sM}{\mathscr{M}}

\newcommand{\cO}{\mathcal{O}}
\newcommand{\cM}{\mathcal{M}}

\newcommand{\bu}{\mathbf{u}}
\newcommand{\bv}{\mathbf{v}}

\newcommand{\derp}{\partial}
\newcommand{\hc}{\text{h.c.}}
\newcommand{\nn}{\nonumber}

\newcommand{\ov}[1]{\overline{#1}}

\newcommand{\unity}{\mathbbm{1}}

\def\Tr{{\texttt{Tr}}}

\def\Re{{\texttt{Re}}}
\def\Im{{\texttt{Im}}}

\def\tH{\widetilde{H}}

\newcommand\subsetsim{\mathrel{%
  \ooalign{\raise0.2ex\hbox{$\subset$}\cr\hidewidth\raise-0.8ex\hbox{\scalebox{0.9}{$\sim$}}\hidewidth\cr}}}

\newcommand{\keV}{\ \text{keV}}

\newcommand{\GeV}{\ \text{GeV}}

\usepackage{color}
\usepackage{colortbl}
\definecolor{rossoc}{cmyk}{0,1,1,0.2}
\definecolor{dgreen}{rgb}{0.0, 0.5, 0.0}

\graphicspath{{Final_Figures/}}

\begin{document} 
\renewcommand*{\thefootnote}{\fnsymbol{footnote}}

\begin{titlepage}

\vspace*{-1cm}
\flushleft{\magenta{IFT-UAM/CSIC-23-65}} 
\\[1cm]
\vskip 1cm

\begin{center}
\blue{\bf \Large The Minimal Massive Majoron Seesaw Model}
\centering
\vskip .3cm
\end{center}
\vskip 0.5  cm
\begin{center}
{\large\bf Arturo de Giorgi}$^{a,b}$~\footnote{\email{arturo.degiorgi@uam.es}},
{\large\bf Luca Merlo}$^{a}$~\footnote{\email{luca.merlo@uam.es}},
{\large\bf Xavier Ponce D\'iaz}$^{c}$~\footnote{\email{xavier.poncediaz@pd.infn.it}},
and \vskip 0.5cm
{\large\bf Stefano Rigolin}$^{c}$~\footnote{\email{stefano.rigolin@pd.infn.it}},
\vskip .7cm
{\footnotesize
$^a$~Departamento de F\'isica Te\'orica and Instituto de F\'isica Te\'orica UAM/CSIC,\\
Universidad Aut\'onoma de Madrid, Cantoblanco, 28049, Madrid, Spain\\
{\par\centering \vskip 0.25 cm\par}
$^b$~Department of Physics \& Laboratory for Particle Physics and Cosmology,\\
Harvard University, Cambridge, MA 02138, USA\\
{\par\centering \vskip 0.25 cm\par}
$^c$~Dipartamento di Fisica e Astronomia ``G.~Galilei" and Istituto Nazionale Fisica Nucleare,\\
Sezione di Padova, Universit\`a degli Studi di Padova, I-35131 Padova, Italy
}

\end{center}
\vskip 2cm
\begin{abstract}
\justify
A convincing explanation of the smallness of neutrino masses is represented by the Type-I Seesaw mechanism, where the two measured neutrino mass differences can be generated by introducing at least two right-handed neutrinos. In an ultraviolet complete model, it is possible to dynamically generate the heavy Majorana scale through the spontaneous symmetry breaking of a global Abelian symmetry and the most economical realisation consists in coupling the two exotic neutral leptons to a singlet complex scalar field. The associated Goldstone boson is often dubbed as Majoron, which may achieve a non-vanishing mass by means of a small term that explicitly breaks the Abelian symmetry. In a generic model, the neutrino and Majoron mass generation mechanisms are completely uncorrelated. In this paper, instead, we reduce the landscape of possible models 
proposing a unique, minimal and predictive framework in which these two types of masses are strictly tied and arise from the same source. 
Bounds from various terrestrial and astrophysical experiments are discussed.
\end{abstract}
\end{titlepage}
\setcounter{footnote}{0}

\pdfbookmark[1]{Table of Contents}{tableofcontents}
\tableofcontents

\renewcommand*{\thefootnote}{\arabic{footnote}}
%
%
%
\section{Introduction}
\label{sec:intro}
%
%

The heterogeneity of the particle masses in the Standard Model (SM) and Beyond (BSM) is one of the big unknowns of modern high-energy physics. No explanation for the large hierarchy of masses and mixings is present within the SM and, at the time being, no convincing evidence of a specific flavour symmetric BSM construction has emerged. In addition, contrary to the SM original ansatz, active neutrinos do have non-vanishing masses and two main frameworks can be introduced to provide them with a mass: Dirac {\it vs.} Majorana. If neutrinos are Dirac fermions, similarly to all the other SM fermions, a right-handed (RH) companion for each flavour is introduced in the spectrum and, preserving the Lepton Number (LN) at tree level, they acquire masses through the ElectroWeak (EW) Spontaneous Symmetry Breaking (SSB) mechanism, proportionally to the Higgs vacuum expectation value (vev), $v_\text{EW}$. This implies, however, that the corresponding Yukawa couplings are tremendously small, deeply worsening the SM flavour puzzle. Conversely, if neutrinos are Majorana fermions, their mass could be associated with the breaking of the LN at a not-well-identified high-energy scale, providing a ``natural'' explanation for their lightness through the well-known Seesaw (SS) mechanism~\cite{Minkowski:1977sc,Gell-Mann:1979vob,Yanagida:1979as,Mohapatra:1979ia}.

An intriguing possibility is that the LN breaking, eventually leading to active neutrino masses at low energy, is a manifestation of high-scale dynamics. Indeed, the Majorana masses can be dynamically generated from the SSB of a global $U(1)$ symmetry at a scale $f_a \gg v_\text{EW}$. Consistently, a Goldstone boson, dubbed as Majoron~\cite{Chikashige:1980qk,
Chikashige:1980ui,Gelmini:1980re}, arises in this context and the $U(1)$ symmetry may be identified with the Peccei-Quinn symmetry associated with the traditional QCD axion framework~\cite{Peccei:1977hh,Weinberg:1977ma,Wilczek:1977pj,
Zhitnitsky:1980tq,Dine:1981rt,Kim:1979if,Shifman:1979if}. Indeed, in its original formulation, the Majoron model has exactly the same ingredients of the KSVZ invisible axion, i.e. a complex scalar field singlet under the SM symmetry and extra exotic heavy fermionic degrees of freedom. The main difference is, of course, the fact that in the Majoron framework, the new exotic fermions are singlets under the SM gauge group, and therefore the Majoron cannot be evoked for solving the strong CP problem. Subsequent works have introduced the Majoron in the context of the Type-II SS~\cite{Choi:1991aa,Chao:2022blc,Biggio:2023gtm}, radiative neutrino models \cite{Bonilla:2023egs,Kannike:2023kzt},
its role as a possible dark matter candidate~\cite{Gelmini:1984pe,Berezinsky:1993fm,Lattanzi:2007ux,Bazzocchi:2008fh,Lattanzi:2013uza,Queiroz:2014yna} and more recently its impact in cosmology has been highlighted~\cite{Ferreira:2018vjj,Arias-Aragon:2020qtn,Arias-Aragon:2020shv,Ferreira:2020bpb,Chun:2023eqc,Chao:2023ojl}, including the possibility that the Majoron may represent a viable solution to the Hubble tension \cite{Escudero:2019gvw,Arias-Aragon:2020qip,Escudero:2021rfi,Araki:2021xdk}.

Differently to what happens for the QCD axion, where the $U(1)_{PQ}$ is explicitly broken by non-perturbative QCD effects thus providing a tiny mass to the axion, the Majoron does not come with an embedded explicit source of symmetry breaking. The mechanism that gives mass to the Majoron has been debated since its formulation~\cite{Chikashige:1980qk,Chikashige:1980ui,Gelmini:1980re}. In Refs.~\cite{Akhmedov:1992hi,Rothstein:1992rh}, for example, it is shown that Planck-suppressed operators explicitly break any global $U(1)$ symmetry, including LN embedded in a continuous Abelian group. Other mechanisms instead involve the active neutrino mass generation mechanism, which includes a LN breaking, in order to equip the Majoron with a mass~\cite{Mohapatra:1982tc,Gu:2010ys,Frigerio:2011in}. In particular, Ref.~\cite{Frigerio:2011in} shows that, in the context of the Type-I seesaw mechanism, a minimal number of terms in the neutral lepton mass matrix is necessary in order to build a massive Majoron model.

In this paper we will extend the results in the literature, introducing a realistic and minimal Seesaw construction 
where LN and PQ are eventually identified and the Majoron mass is strictly linked to the active neutrino masses. 
The starting point of our analysis is the minimal Type-I SS, where the SM fermion spectrum is enlarged by only two 
RH neutrinos, also called Heavy Neutral Leptons (HNLs), that are singlets under the whole SM group. 
In this construction, therefore, at most two active neutrinos can become massive, while the lightest one remains massless. 
It is possible, then, to embed a dynamical $U(1)_\text{PQ}$ SSB mechanism, the Majoron being the associated Goldstone boson. 
We will work on this setup by adopting the following “minimality” requirements: 
\begin{description}
\item[i)] only renormalisable interactions are considered in the Lagrangian densities;
\item[ii)] only one large (Majorana) scale is present, and associated with the SSB of the $U(1)_\text{PQ}$ symmetry, i.e. only one complex scalar field couples to the RH neutrinos;
\item[iii)] only one (small) explicit $U(1)_\text{PQ}$-violating term is introduced.
\end{description}
We will show that satisfying these three conditions will lead to a unique and predictive model where the active neutrino masses and the 
Majoron mass are deeply connected, i.e. the explicit symmetry-violating term is the necessary and sufficient ingredient to 
simultaneously provide a mass to the active neutrinos and the Majoron.

The paper is structured as follows. In Sect.~\ref{sec:Lagrangian}, we briefly review the generation of active neutrino 
masses in different SS realisations with only two RH neutrinos, first with the traditional type-I and then in Sect.~\ref{sec:lowscale} the low-scale SS. In Sect.~\ref{sec:LSSRadiative}, we explicitly derive the active neutrino mass matrix including one-loop contributions induced by a heavy but non-degenerate pair of HNLs. 
Sect.~\ref{sec:mmM} describes the minimal massive Majoron Seesaw (mmM) model. We 
explicitly extend the SM Lagrangian by specific couplings between the HNLs and a complex scalar field, $\phi$, singlet under 
the SM group, but endowed with a $U(1)_{\textrm{PQ}}$ symmetry. Once this scalar acquires a vev, $f_a$, the PQ gets spontaneously 
broken giving rise to the Majoron. In Sect.~\ref{sec:UVCompletion}, we analyse different classes of ultraviolet (UV) 
embeddings that can be constructed with one complex scalar and two RH neutrinos. In Sect.~\ref{sec:mmMlag} we 
identify the mmM model that respects the three minimality conditions i)-iii). Then, we determine the parameter space in which 
higher-order contributions to the neutrino mass matrix do not spoil perturbativity, and therefore a realistic active neutrino 
spectrum and PMNS mixing can be predicted.

Sect.~\ref{sec:MajoronMass} represents the core of this paper, containing the one-loop derivation of the Majoron mass. A preliminary calculation is 
shown for a simplified model in Sect.~\ref{sec:toymodel}, both in the chirality flipping and chirality preserving basis 
for the Majoron, and then in Sect.~\ref{sec:radiative} the Majoron mass in the mmM model is derived, in the chirality 
preserving basis where the calculations are greatly simplified. Finally, in Sect.~\ref{sec:Pheno}, we discuss the possible 
phenomenology of this Majoron and compare it with the present bounds under the context of general ALP and Dark Matter 
searches. This work is completed with three Appendices: App.~\ref{app:MassEigen} deals with the diagonalisation of neutrino 
mass matrix and Majoron interactions; App.~\ref{app:Majoron-interactions} describes the HNL interactions with the Majoron 
and SM gauge and Higgs bosons; and finally, the one-loop Coleman-Weinberg (CW) potential~\cite{Coleman:1973jx} is derived 
in App.~\ref{app:CW}, as a crosscheck of the diagrammatic calculation performed in Sect.~\ref{sec:MajoronMass}. 

%
%
\section{Seesaws with two HNLs}
\label{sec:Lagrangian}
%
%

The Type-I SS mechanism~\cite{Minkowski:1977sc,Gell-Mann:1979vob,Yanagida:1979as,Mohapatra:1979ia} provides a natural 
explanation of the smallness of the active neutrino masses by introducing RH neutrinos. To be able to explain the two 
neutrino oscillation mass differences at least two HNLs, singlets under the whole SM group, have to be introduced. 
Therefore, at most two active neutrinos can become massive while the lightest one remains massless. In a compact 
notation, the SM and exotic neutral lepton fields can be grouped in a left-handed lepton multiplet denoted as,
\be
\chi_L\equiv(\nu_L,\,N_R^c,S^c_R)^T\,,
\label{NeutralVector}
\ee
where $\nu_L \equiv (\nu^e_L, \nu^\mu_L,\nu^\tau_L)$ are the SM neutrinos and $N_R$ and $S_R$ are the two HNLs, 
whose conjugates are defined as $\psi^c_R = \mathcal{C} \overline{\psi}_R^T$, where $\mathcal{C} $ is the charge conjugation matrix.

In this scenario, the most general (renormalizable) Lagrangian describing the neutral lepton interactions reads:
\be
\begin{split}
-\sL_\text{LN} = & \phantom{+} \ov{L_L}\,\tH\, Y_N\,  N_R + \ov{L_L}\,\tH\,Y_S\,S_R+\\
&+\dfrac12 \Big[ \Lambda_{NN}\,\ov{N_R^c}\,N_R + \Lambda_{SS}\,\ov{S_R^{c}}\,S_R+ \Lambda_{NS}\,(\ov{N_R^{c}}\,S_R + 
\ov{S_R^{c}}\,N_R) \Big]+ \hc \, .
\end{split}
\label{eq:TypeILN}
\ee
where $L_L$ is the EW lepton doublet, triplet in flavour space, and $H$ is the Higgs EW doublet, with $\tH\equiv 
i\sigma_2 H^\ast$. $Y_{N,S}$ are two generic three-dimensional vectors describing the Dirac-type Yukawa interactions 
with the Higgs, while $\Lambda_{NN}$, $\Lambda_{SS}$ and $\Lambda_{NS}$ are three one-dimensional parameters. 

After EW SSB, with the Higgs developing a vev $v_\text{EW}=246\GeV$, the following neutral lepton mass matrix 
is generated
\be
-\sL_\text{$\nu$M} \supset \dfrac{1}{2}\,\ov{\chi}_L\, \mathcal{M}_\chi \, \chi_L^c 
\qquad \mbox{with} \qquad \mathcal{M}_\chi =
\begin{pmatrix}
	0 & m_N & m_S \\[2mm]
	m_N^T & \Lambda_{NN} & \Lambda_{NS}\\[2mm]
	m_S^T & \Lambda_{NS} & \Lambda_{SS} \\
\end{pmatrix}  \equiv 
\begin{pmatrix}
	0 & \widehat{m} \\[2mm]
	\widehat{m}^T & \widehat{\Lambda} \\
\end{pmatrix} \,,
\label{GenericNeutralMassLag}
\ee
where the Dirac mass terms are defined as $m_{N,S}\equiv Y_{N,S}\, v_\text{EW} /\sqrt{2}$. Sometimes it will 
be useful to use the compact notation $\widehat{m}$ and $\widehat{\Lambda}$ for indicating the $3\times 2$ and 
$2\times 2$ Dirac and Majorana mass terms. For example, diagonalising $\mathcal{M}_\chi$, a mass term for the 
active neutrinos appears and the corresponding mass matrix is given by
\be
m_\nu^\text{Type-I}\simeq - \widehat{m}\,\widehat{\Lambda}^{-1}\widehat{m}^T\,.
\label{ActiveNuMassTypeI}
\ee

The values of the Dirac and Majorana masses are fixed in order to reproduce the active neutrino masses and the PMNS mixings. 
Assuming no large hierarchies within the entries of the Dirac Yukawas, $(\widehat{m})_{ij}=\mathcal{O}(v_\text{EW})$, 
in order to reproduce the neutrino mass squared differences, one is forced to take the overall scale of the Majorana mass 
matrix as $\Tr\,\widehat\Lambda \sim 10^{14}-10^{15} \GeV$. The latter is approximately the mass of the HNLs, after the 
diagonalisation of $\mathcal{M}_\chi$, and therefore it is practically impossible to observe any effect of the HNLs at 
present/future colliders or flavour factories.

In the original construction of the Type-I SS mechanism, all the leptons have the same transformation properties under LN 
and customarily $L(L_L)=1=L(N_R)=L(S_R)$ is chosen. It follows that the Dirac terms in Eq.~\eqref{eq:TypeILN} are LN invariant, 
while the Majorana mass terms violate LN by two units. A useful exercise consists of interpreting the Dirac Yukawas and 
the Majorana masses as spurion fields, that is non-dynamical fields that may own transformation properties, in this specific 
case, only under LN. Hence,  one can formally implement LN invariance of the whole Lagrangian, assigning specific charges 
to the spurions: whenever a spurion charge is different from zero, the corresponding term would violate LN. Applying this 
spurionic description to the Lagrangian in Eq.~\eqref{eq:TypeILN}, one obtains $L(\widehat{m})=0$ and $L(\widehat\Lambda)=-2$, 
confirming that the Majorana terms violate LN.  Moreover, we obtain that the active neutrino mass matrix also violates 
LN as $L(m_\nu^\text{Type-I})=2$, as expected. 

Going beyond the original setup, modifying the LN charge assignments would change the previous conclusions, without necessarily 
affecting the physical observables. For example, if we fix $L(L_L)=1$ and $L(N_R)=L(S_R)=n\neq1$, the Dirac terms would now 
violate the LN as indeed $L(\widehat{m})=1-n$, while the Majorana ones may or may not violate it, depending on the explicit 
value of $n$, being $L(\widehat\Lambda)=-2n$. However, the active neutrino mass matrix still violates LN by the same quantity 
as in the traditional case, $L(m_\nu^\text{Type-I})=2$, for any value of $n$. 

This simple exercise points out the following:
\begin{itemize}
\item[-] There are several LN charge assignments that lead to the same physics.
\item[-] The LN conserving or violating status of a Lagrangian term can only be assessed with respect to a specific choice 
of the charges. Moreover, we can conclude that LN conservation can only occur if there is a charge assignment such that all 
the LN spurions, $\widehat{m}$ and $\widehat\Lambda$ in this case, have vanishing charges.
\item[-] The active neutrino masses necessarily depend on (a combination of) the spurions that, in any charge assignment, 
have a non-zero charge: this is to say that LN is broken by the simultaneous presence of these spurions, as otherwise, if any 
of them vanishes, the active neutrinos would remain massless. For the Type-I SS case, as there exist charge assignments such 
that both $\widehat{m}$ and $\widehat\Lambda$ are explicitly LN violating implies that both quantities have to appear in the 
definition of $m_\nu^\text{Type-I}$, confirming the result of the explicit computation and that both of them should be 
non-vanishing to assure massive active neutrinos. However, in more complicated setups with respect to the Type-I SS, it may 
occur that LN-violating spurions appear only in loop-level contributions to the active neutrino masses. Indeed, a broken 
symmetry does not necessarily imply that its effects are manifest in observables described with tree-level Feynman diagrams.
\end{itemize}

We will use and adapt this reasoning in the next sections, where we will go through other popular SS mechanisms, and show 
that the very last condition helps identify the genuine LN-violating spurions.

%
\subsection{The Low-Scale Seesaw models}
\label{sec:lowscale}
%

A popular modification of the canonical Type-I Seesaw is the class of constructions that undergo the name of low-scale Seesaw (LSSS) 
mechanisms \cite{Branco:1988ex,Kersten:2007vk,Abada:2007ux,Moffat:2017feq}, also known as ``LN protected'' SS mechanisms. 
In this kind of scenarios, the two HNLs have different non-vanishing LN charge assignments. For example, assuming 
$L(L_L) = L(N_R)=-L(S_R) = 1$ leads to the following LN conserving Lagrangian:
\be
-\sL_\text{LN}=\ov{L_L}\,\tH\, Y_N\,  N_R + \frac{\Lambda_{NS}}{2}\,
\left(\ov{N_R^{c}}\, S_R +\ov{S_R^{c}}\, N_R \right) + \hc
\label{eq:TypeIbLNC}
\ee
while (a combination of) additional terms, explicitly violating the LN symmetry, have to be added to generate the desired 
light active neutrino spectrum: in all generality,
\be 
-\sL_{\epsilon \text{LN}} = \ov{L_L}\,\tH\, \epsilon Y_S\,  S_R + \dfrac{\epsilon \Lambda_{NN}}{2}\,\ov{N_R^c}\,N_R + 
\dfrac{\epsilon \Lambda_{SS}}{2}\, \ov{S_R^{c}}\, S_R + \hc \,.
\label{eq:TypeIbLNV}
\ee
By the physical assumption of an approximate LN symmetry, the three parameters $\epsilon Y_S,\epsilon \Lambda_{NN}$ and 
$\epsilon \Lambda_{SS}$ are naturally small compared to the LN preserving ones appearing in Eq.~\eqref{eq:TypeIbLNC}.

In the broken EW phase, one obtains the following neutrino mass matrix 
\be
-\sL^\text{LSSS}_\text{$\nu$M}\supset\dfrac{1}{2}\,\ov{\chi}_L\cM^\text{LSSS}_\chi \chi_L^c \qquad\text{with}\qquad
\cM^\text{LSSS}_\chi=
\begin{pmatrix}
	0 & m_N & \epsilon m_S\\[2mm]
	m_N^T & \mu' & \Lambda_{NS} \\[2mm]
	\epsilon m^T_S & \Lambda_{NS} & \mu \\
\end{pmatrix}\,,
\label{NeutralMassLagLSS}
\ee
with the obvious definitions of the Dirac mass terms, inherited from the Lagrangian densities in Eq.~\eqref{eq:TypeIbLNC} and 
\eqref{eq:TypeIbLNV}, and where $\mu'=\epsilon \Lambda_{NN}$ and $\mu=\epsilon \Lambda_{SS}$ have been used to make contact 
with the notation often used in the literature.  Adopting the spurion description, with the chosen charge assignment, the 
parameters of the LN violating terms should be promoted to spurions with non-vanishing charges, $L(\epsilon m_S,\mu',\mu)
=(+2,-2,+2)$, while the quantities $m_N$ and $\Lambda_{NS}$ do not acquire any charge.

After the EW SSB, at leading order in the $\mu^{(\prime)}/\Lambda_{NS}$ and $\epsilon m_S/\Lambda_{NS}$ expansion, the active 
neutrinos mass matrix is given by 
\be
m_\nu^\text{LSSS}\simeq - \mu \, \dfrac{m_N m_N^T}{\Lambda_{NS}^2}-\epsilon\dfrac{m_S\,m_N^T + m_N\,m_S^T}{\Lambda_{NS}}\,.
\label{genericLSSmnu}
\ee
Having introduced only two exotic HNLs, this implies again that the lightest active neutrino remains massless. The other two neutrinos 
acquire masses that are functions of $\epsilon m_S$ and $\mu$, but not of $\mu'$ that does not play any role at leading order. 

The advantage of this class of LSSS models is that having introduced ``naturally'' small terms, i.e. $\epsilon m_s \ll m_N$ and 
$\mu \ll \Lambda_{NS}$, one can explain the active neutrino masses by introducing a much lighter Majorana scale. To reproduce 
the observed neutrino mass spectrum it is sufficient to fix $\epsilon m_S (\mu) \sim 10 (1000)$ eV for a chosen $\Lambda_{NS} 
\sim\cO(\mbox{TeV})$. It follows that, while active neutrino masses remain small, the HNLs are relatively light and possibly 
detectable at colliders. Moreover, the unique $d=6$ effective operator resulting from integrating out the HNLs does not depend 
on the LN violating parameters, thus describing possibly interesting phenomenological effects in both direct and indirect 
searches~\cite{Abada:2007ux}. Although TeV scale HNLs is a very attractive feature of these constructions, it has to be 
pointed out that the texture in Eq.~\eqref{NeutralMassLagLSS} provides a good description of the low-energy neutrino data 
even for larger $\Lambda_{NS}$ by accordingly rescaling the LN violating parameters. 

Two popular models in this scenario are the ones obtained by setting $\epsilon m_S=\mu'=0$, dubbed as Inverse Seesaw 
(ISS)~\cite{Akhmedov:1995ip,Malinsky:2005bi} or by imposing $\mu=\mu'=0$, dubbed as Linear Seesaw (LSS)~\cite{Mohapatra:1986aw,
Mohapatra:1986bd}, that predict the following active neutrino masses respectively
\be
m_\nu^\text{ISS}\simeq - \mu \, \dfrac{m_N m_N^T}{\Lambda_{NS}^2}\,,  \qquad\qquad 
m_\nu^\text{LSS}\simeq - \epsilon\dfrac{ m_S\,m_N^T + m_N\, m_S^T}{\Lambda_{NS}}\,.
\ee
Notice that in the ISS case, it is not possible to describe successfully the neutrino spectrum and the PMNS mixing matrix 
with only two HNLs, as the product $m_N m_N^T$ has rank one. On the other hand, in the Linear Seesaw case, the light neutrino 
mass matrix has, instead, rank 2 and allows for a description of the neutrino sector compatible with data as discussed in 
Ref.~\cite{Gavela:2009cd}.

Before concluding this section, we generalise the charge assignment and discuss the spurion role as we did for the Type-I SS case. 
First of all, one key hypothesis of this setup is that $L(N_R)=- L(S_R)\neq0$ and as a result the term associated to $\Lambda_{NS}$ 
is automatically LN invariant, while the two Majorana terms proportional to $\mu$ and $\mu'$ always violate LN, as $L(\mu)=-2L(N_R)$ 
and $L(\mu')=-2L(S_R)$. To provide a mass for the active neutrinos, the charge of the lepton doublet should be fixed such that 
$L(L_L)=L(N_R)$: this guarantees that the term proportional to $Y_N$ is invariant under LN.\footnote{The alternative choice 
$L(L_L)=L(S_R)$ leads to exactly the same physics, as indeed it is sufficient to exchange $N_R$ with $S_R$ to obtain the same 
neutral lepton mass matrix.} All in all, a non-vanishing active neutrino mass would therefore involve $\mu$ and $\mu'$, although 
not necessarily the two at the same time. The ISS case is the example in which only the term proportional to $\mu$ is switched 
on as an explicit breaking of LN and a tree-level active neutrino mass is generated. On the other hand, although we may expect 
the same for $\mu'$, this is not the case: no tree-level contributions to the active neutrino masses arise in this case, but they 
arise at one-loop~\cite{Pilaftsis:1991ug}. Focusing now on the Dirac terms, if we fix $L(L_L)=1$ and $L(N_R)=-L(S_R)=n\neq\pm1$, 
we conclude that both $Y_N$ and $Y_S$ should acquire a charge, $L(Y_N)=1-n$ and $L(Y_S)=1+n$, thus breaking LN. Notice that this 
is independent of the presence of the terms proportional to $\mu$ and $\mu'$. This implies that the active neutrino masses 
necessarily contain the product of the two Dirac Yukawas, consistently with the explicit computation in the LSS mechanism.

An interesting alternative option with respect to the Inverse and Linear SS mechanisms is when $\mu=0$, but both $Y_S$ and 
$\mu'$ are added to the Lagrangian. As we will discuss in the next section, this may have a deep impact in the active neutrino 
mass generation and, moreover, represents the optimal setup for a massive Majoron.

%
%
\subsection{The Extended Seesaw limit}
\label{sec:LSSRadiative}
%

In this section, we further exploit the spurionic approach in a modified LSS Lagrangian, which has been named as Extended Seesaw 
limit in Ref.~\cite{Lopez-Pavon:2012yda}, whose physics case will be worked out in the next section. Using the same notation of 
Sect.~\ref{sec:lowscale}, we separate the neutral leptonic Lagrangian in a part that is invariant under LN and a part that 
explicitly violates it, once fixing for definiteness the LN charges as $L(L_L) = L(N_R) = 1$, $L(S_R)=-3$, $L(\Lambda_{NN})=-2$ 
and $L(\Lambda_{NS})=2$:
\begin{align}
	\label{eq:neutrinosector}
	-\sL_\text{LN} = & \ov{L_L}\,\tH\, Y_N\, N_R + \dfrac{\Lambda_{NN}}{2}\,\ov{N_R^c}\,N_R + 
	\dfrac{\Lambda_{NS}}{2}\,(\ov{N_R^{c}}\, S_R + \ov{S_R^{c}}\, N_R) + \hc \\
	\label{eq:neutrinosectorV}
	-\sL_{\epsilon LN} = & \epsilon\ov{L_L}\,\tH\,Y_S\,S_R + \hc \,. 
\end{align}
The only term that explicitly breaks LN is the one proportional to $\epsilon Y_S$ that is assumed to be $\epsilon Y_S \ll Y_N$. 
On the other hand, once $\Lambda_{NN}$ and $\Lambda_{NS}$ acquire a background value, they represent two large Majorana masses 
such that $\Lambda_{NN} \sim \Lambda_{NS} \gg v_\text{EW}$.

After SSB of the EW symmetry, the corresponding neutral lepton mass matrix reads:
\be
-\sL^\text{ESS}_\text{$\nu$M}\supset \dfrac{1}{2}\ov{\chi_L}\,\cM^{\text{ESS}}_\chi\chi_L^c+\hc\qquad\text{with}\qquad
\cM^{\text{ESS}}_\chi=\begin{pmatrix}
	0 &  m_N & \epsilon\, m_S \\
	m_N^T & \Lambda_{NN} & \Lambda_{NS} \\
	\epsilon\, m_S^T & \Lambda_{NS} & 0
\end{pmatrix}\,,
\label{MajoronNeutralMassMatrix}
\ee
where $\Lambda_{NN}$ plays the role that in the traditional ISS mechanism belongs to $\mu'$, although the similarities end here, 
as there is a tree-level contribution to the active neutrino masses proportional to $\Lambda_{NN}$. Indeed, the active neutrino 
mass matrix at tree level is given by
\be
m^\text{TL}_\nu=-\epsilon \dfrac{ m_S\,m_N^T+m_N\,m_S^T}{\Lambda_{NS}} + 
\epsilon^2 \, \frac{\Lambda_{NN}}{\Lambda_{NS}} \, \dfrac{m_S \, m_S^T}{\Lambda_{NS}}\,,
\label{ActiveNuMass}
\ee
although the second term is typically negligible as $\epsilon \ll 1$ is assumed. On the other hand, this condition translates 
to an upper bound for the Majorana scale: for example, $\epsilon < 10^{-3}$ implies that $\Lambda_{NS} < 10^{12}\GeV$, to 
reproduce the observed atmospheric mass splitting.

It is well known in the literature~\cite{Pilaftsis:1991ug,Grimus:2002nk,AristizabalSierra:2011mn,Lopez-Pavon:2012yda}, that 
large values of $\Lambda_{NN}$ and $\Lambda_{NS}$ can generate a sizable mass splitting between the two HNL masses, implying 
non-negligible and possibly dangerous one-loop corrections to the tree-level result in Eq.~\eqref{ActiveNuMass}. The calculation 
and the discussion of the phenomenological consequences of such loop corrections will be detailed in the following subsection.

\subsubsection{Active neutrino masses at one-loop}

Let us start by diagonalising the HNL sector of the mass matrix in Eq.~\eqref{MajoronNeutralMassMatrix}. The leading contribution, 
in the $v_{EW}/\Lambda$ expansion, to the masses of the HNL states reads:
\be
M_{N,S}=\dfrac{\Lambda_{NS}}{2}\left[\sqrt{4+\left(\dfrac{\Lambda_{NN}}{\Lambda_{NS}}\right)^2}\mp 
\left(\dfrac{\Lambda_{NN}}{\Lambda_{NS}}\right)\right]\,.
\label{eq:HNL-masses}
\ee
Notice that positively defined masses can be obtained by redefining the lightest eigenvector with a Majorana phase $i$.

The active neutrino mass matrix receives one-loop contributions from diagrams involving the HNLs and either the Higgs 
or the $Z$ gauge boson. At leading order in $v_{EW}/\Lambda$ one obtains~\cite{Pilaftsis:1991ug,Grimus:2002nk,
AristizabalSierra:2011mn,Lopez-Pavon:2012yda}:
\be
\label{eq:numassloop}
\delta m_\nu^\text{1L}=2\,\dfrac{m_N\, m_N^T}{(4\pi v)^2}\frac{M_H^2+3M_Z^2}{M_{N}+M_{S}}\log{\left(\dfrac{M_S}{M_N}\right)}\,.
\ee
While in the degenerate limit, $M_S \sim M_N$, i.e. $\Lambda_{NN} \sim 0$, the Higgs and $Z$ boson contributions cancel 
each other, when $\Lambda_{NN}\sim \Lambda_{NS}$ the HNLs mass splitting induces sizable one-loop corrections. Summing 
together the tree- and one-loop contributions, the neutrino mass matrix can be written as

\begin{equation}
m_\nu = m^\text{TL}_\nu+\delta m_\nu^\text{1L}\equiv m_{T_1} (\bu \bv^T+\bv \bu^T)+m_{T_2}\bv\bv^T+m_L \bu\bu^T\,,
\label{numassLoop}
\end{equation}
where the vectors $\bu$ and $\bv$ 
\be
\bu\equiv  \dfrac{Y_N}{|Y_N|} \qquad \,, \qquad \bv\equiv\dfrac{Y_S}{|Y_S|}\,, 
\ee
define the directions in the flavour space and where the tree-level and one-loop overall numerical contributions 
$m_{T_{1}},m_{T_{2}}$ and $m_{L}$ in terms of the Dirac Yukawas and Majorana mass terms read
\bea
m_{T_1} & = & -\epsilon \dfrac{|Y_N| |Y_S| v_\text{EW}^2}{2\Lambda_{NS}} \qquad \,, \qquad 
m_{T_2}=\epsilon^2 \, \frac{\Lambda_{NN}}{\Lambda_{NS}} \, \dfrac{|Y_S|^2v_\text{EW}^2}{2\Lambda_{NS}}\,,
\label{eq:mT1} \\
m_L & = & \dfrac{|Y_N|^2}{16\pi^2}\dfrac{M_H^2+3M_Z^2}{M_{N}+M_{S}}\log{\left(\dfrac{M_S}{M_N}\right)}\,.
\label{eq:ma1}
\eea

To obtain the analytic expressions for the light neutrino mass eigenvalues, we generalise and adapt the procedure introduced 
in Ref.~\cite{Gavela:2009cd}, including the extra tree-level and one-loop contributions. The two non-zero eigenvalues read
\be
\begin{split}
    \label{eq:mpm}
    |m_\pm|^2=\dfrac{1}{2}&\left[m_C^2-\tau^2(2m_{T_1}^2-m_{T_2}^2-m_{L}^2)\right.\\
    &\qquad\left.\pm\sqrt{\left(m_C^2-\tau^2(2m_{T_1}^2-m_{T_2}^2-m_{L}^2)\right)^2-4\tau^4(m_{T_1}^2-m_L m_{T_2})^2}\right]\,,
\end{split}
\ee
where 
\begin{align}
    &\label{etaDEF} m_C\equiv |2m_{T_1}+\eta m_L+\eta^\ast m_{T_2}|\,,&&\eta\equiv \bu^\dagger \bv\equiv |\eta|e^{i\vartheta_\eta}\,, 
    && \tau^2 \equiv 1-|\eta|^2\,.
\end{align}

Conventionally, we have chosen $m_{T_1}<0$ and $\theta_\eta\in[-\pi/2,\pi/2]$. This has no impact on the results as the sign of $m_{T_1}$ 
can be adjusted in $m_C$ by a shift of $\pi$ in the phase $\vartheta_\eta$. Contrary to the case in Ref.~\cite{Gavela:2009cd}, the phase 
$\vartheta_\eta$ is a physical parameter to the presence of the extra tree- and loop-contributions: only neglecting $m_{T_2,L}$ it is 
possible to redefine away $\vartheta_\eta$.~\footnote{Notice that, introducing a Majoron in this construction, even if $m_{T_2,L}\neq 0$, 
the Majoron potential dynamically relaxes such parameter to $\theta_\eta=0$, as discussed in Appendix \ref{app:CW}.} The vectors \textbf{$u$} and \textbf{$v$} are fixed to reproduce the PMNS mixing angles, following Ref.~\cite{Gavela:2009cd}. As there are enough free parameters to correctly account for the observed values, they do not impose any constraints on the model\footnote{They enter, however, 
in fixing the flavour structure of the Majoron couplings, see Sec.~\ref{sec:Pheno}.} For the Normal Ordered 
(NO) and Inverted Ordered (IO) spectra, we thus have
\begin{align}
&\text{NO:} &&|m_1|^2 = 0\,, &&|m_2|^2 = |m_-|^2\,, &&|m_3|^2 = |m_+|^2\,,\\
&\text{IO:} &&|m_1|^2 = |m_-|^2\,,  &&|m_2|^2 = |m_+|^2\,,
&&|m_3|^2 = 0\,.
\end{align}
We further define the ratio of the solar and atmospheric neutrino mass-splittings as in Ref.~\cite{Esteban:2020cvm}
\begin{equation}
r\equiv \dfrac{|\Delta m_\text{sol.}^2|}{|\Delta m_\text{atm.}^2|}\equiv\begin{cases}
\dfrac{|\Delta m_{21}^2|}{|\Delta m_{31}^2|}=\dfrac{|m_-|^2}{|m_+|^2}\,,  &\qquad\text{for the NO}\,, \\
& \\
\dfrac{|\Delta m_{21}^2|}{|\Delta m_{32}^2|}=\dfrac{|m_+|^2-|m_-|^2}{|m_+|^2}\,,&\qquad\text{for the IO}\,.
\end{cases}
\end{equation}

In the limit $m_{T_2,L}\ll m_{T_1}$, the expression in Eq.~\eqref{eq:mpm} greatly simplifies and reads
\be
\label{eq:mplusmminus}
|m_\pm|^2\approx m_{T_1}^2\left(1\pm|\eta|\right)^2\,,
\ee
which matches with the result of Ref.~\cite{Gavela:2009cd}. Along the same lines, it is convenient to extract the value of $|\eta|$ 
from the neutrino mass difference ratio:
\begin{align}
&|\eta|\equiv \dfrac{1-\sqrt{r}}{1+\sqrt{r}}\,,&&\text{for the NO}\,,
\label{eq:NOdef}\\
&|\eta|\equiv \dfrac{1-\sqrt{1-r}}{1+\sqrt{1-r}}\,,&&\text{for the IO}\,.
\label{eq:IOdef}
\end{align}
These approximate expressions are valid given that, as it will turn out in our UV completed model, the contributions of $m_{T_2,L}$ 
are at most $\mathcal{O}(10\%)$ in the considered parameter space. Therefore, these will be the formulae used from now on.

%
%
%
%
\section{The minimal massive Majoron Seesaw model}
\label{sec:mmM}
%

All the discussion in the previous section concerns the LN symmetry of the infrared (IR) theory. However, a theoretically intriguing assumption is that the Majorana mass terms in the low-energy Lagrangian of Eq.~\eqref{eq:TypeILN} have a dynamical origin in the 
UV-theory, through the SSB of the $U(1)_\text{PQ}$ at some scale $f_a \sim \Lambda \gg v_\text{EW}$, similarly to what happens in the 
SM where Dirac mass terms arise from the SSB of the $SU(2)_L$ symmetry. In the following, we discuss the embedding of a dynamical 
SSB mechanism in the models with two HNLs described in the previous section, respecting three ``minimality'' requirements: 
i) only renormalisable interactions are considered in the Lagrangian densities; ii) only one large (Majorana) scale is present 
and is associated with the SSB of the PQ symmetry, i.e. only one complex scalar field couples to the HNLs; and iii) only one (small) 
explicit LN and PQ-violating term is introduced. In particular, we assume that there is no other PQ symmetry-breaking term than the one 
in the Yukawa sector. We will see that, once satisfied these three conditions, a unique model that correctly describes the active neutrino 
spectrum and PMNS mixing also predicts a tight correlation between these masses (i.e. associated with the LN breaking) and the Majoron 
mass (i.e. resulting from the PQ breaking).

According to the previous criteria, the SM scalar spectrum is extended only by a single complex scalar field $\phi$, endowed with 
a $U(1)_\text{PQ}$ global symmetry, that gets spontaneously broken by its non-vanishing vev, $f_a$. It is customary to define
\be
\phi\equiv \dfrac{(f_a + \rho)}{\sqrt2} e^{ia/f_a}\,,
\label{phiDEF}
\ee 
being $\rho$ the radial mode and $a$ the GB associated with the SSB of the PQ symmetry, hereafter dubbed as Majoron. The scale 
$f_a$ is assumed to be much larger than the EW scale, $f_a\gg v_\text{EW}$, in such a way that the radial field can be integrated 
out and the Majoron remains the only scalar light degree of freedom at low energies, besides the Higgs. 

Although the RH neutrinos are gauge singlets, they can, in general, transform both under the LN and PQ symmetries and therefore 
can couple both with the SM leptons (and the Higgs) and the scalar field $\phi$. Therefore, when the PQ symmetry gets spontaneously 
broken the Majorana mass terms for the RH neutrinos are dynamically generated. The manifestation of the LN breaking in the neutral 
lepton mass matrix of Eq.~\eqref{GenericNeutralMassLag} can be easily traced by introducing LN spurionic charges, as already 
illustrated in the previous section. Thus, to identify the specific LN-violating pattern, it is sufficient to impose a specific 
LN charge assignment to the neutral leptons and read the spurionic charges of the terms in the neutral lepton mass matrix. Giving 
a vanishing LN charge to the Higgs simplifies the exercise. 

On the other hand, a different reasoning is necessary to identify the possible sources of the PQ symmetry breaking, as we will 
discuss in the next section. In principle one could also consider additional PQ-violating terms in the scalar potential, at 
the cost of washing out the Majoron-neutrino mass correlation studied in this paper, contrary to the minimality condition assumed 
from the beginning.\\

%
%
\subsection{Spurionic analysis of the PQ symmetry}
\label{sec:UVCompletion}

It is relatively easy to introduce the field $\phi$ in the SS realisations discussed in the previous section, giving rise to 
SSB of the LN. Accordingly to the minimality condition i) of having a renormalisable Lagrangian, we can write
\be
\begin{split}
-\sL_\text{PQ} = & \phantom{+}\ov{L_L}\,\tH\, Y_N\,  N_R + \ov{L_L}\,\tH\,Y_S\,S_R+ \\ 
& + \frac{1}{2}\,\phi\,\Big[Y_{NN}\,\ov{N_R^c}\,N_R + Y_{SS}\,\ov{S_R^{c}}\,S_R + Y_{NS}\, 
\left(\ov{N_R^{c}}\,S_R + \ov{S_R^{c}}\,N_R\right) \Big] + \hc \,,
\end{split}
\label{eq:TypeILNC}
\ee
where the Dirac, $Y_{N,S}$, and Majorana, $Y_{NN,NS,SS}$, Yukawa terms are large or small depending on the underlying 
LN symmetry assumed for each scenario. For example, the Linear Seesaw case is obtained, after the EW and PQ SSB, for
\be
Y_S\rightarrow \epsilon Y_{S}\,,\qquad\qquad 
Y_{NN}=0=Y_{SS}
\ee
and identifying
\begin{equation}
m_{N}= \frac{Y_{N}}{\sqrt{2}} v_\text{EW}\,, \qquad \qquad \epsilon m_{S}= \frac{\epsilon Y_{S}}{\sqrt{2}} v_\text{EW} \,, 
\qquad \qquad \Lambda_{NS} = \frac{Y_{NS}}{\sqrt{2}} f_a \,,
\label{DEFmM1}
\end{equation}
with the ``natural'' hierarchy, from the LN charge assignment point of view, $\epsilon Y_S \ll Y_N$. 

It is straightforward to observe that, besides the LN, the Lagrangian in Eq.~\eqref{eq:TypeILNC} possess an unbroken 
$U(1)_\text{PQ}$ symmetry, with charge assignment
\be
PQ(L_L)=PQ(N_R)=PQ(S_R)=-PQ(\phi)/2\,.
\label{PQChargeAssignmentTypeILNCV}
\ee
As a consequence, the Majoron originated within the PQ SSB remains massless. This can be explicitly seen performing 
the following field redefinitions:
\be
\chi_L \rightarrow e^{-\frac{i}{2}\frac{a}{f_a}} \chi_L\,,
\label{PQredefinition}
\ee
that remove the Majoron dependence in all the Yukawa terms\footnote{The simultaneous redefinition of the RH charged 
lepton fields, $e_R\rightarrow e^{-\frac{i}{2}\frac{a}{f_a}} e_R$, ensures the removal of the Majoron dependence in 
the charged lepton Yukawa interaction. Being a vectorial transformation implies no generation of anomalous 
couplings with the gauge bosons.}. The Majoron dependence reappears then in the Lagrangian through the kinetic terms 
as derivative interactions, signalling the underlying presence of the GB shift symmetry.  

In order to give a mass to the Majoron, small PQ-violating terms in the Lagrangian can be introduced:
\bea 
-\sL_{\epsilon \text{PQ}} & = & \frac{1}{2}\, \phi^* \,\Big[\epsilon Y_{NN}\,\ov{N_R^c}\,N_R + \epsilon Y_{SS}\,\ov{S_R^{c}}\,S_R + 
\epsilon Y_{NS}\, \left(\ov{N_R^{c}}\,S_R + \ov{S_R^{c}}\,N_R\right) \Big]\,,
\label{eq:TypeILNV}
\eea
being $\epsilon Y$ naturally tiny parameters ``protected'' by the PQ symmetry. It is impossible now to simultaneously 
eliminate the Majoron dependence from both the Yukawa Lagrangian in Eq.~\eqref{eq:TypeILNC} and that in Eq.~\eqref{eq:TypeILNV} 
by a field redefinition alike in Eq.~\eqref{PQredefinition}, implying the presence of shift symmetry violating terms 
in the theory and consequently a (loop generated) mass for the Majoron. 

The additional terms introduced in Eqs.~\eqref{eq:TypeILNV} also contribute to the active neutrino masses, but clearly 
only through sub-dominant effects with respect to the leading contributions in Eq.~\eqref{eq:TypeILNC}. However, this 
implies that the Majoron mass is not correlated to the active neutrino masses, or said otherwise the explicit PQ breaking 
can be considered as an {\it ad hoc} ingredient to provide the Majoron with a mass --equivalent to an explicit Majoron 
mass term in the scalar potential-- rather than being a common feature of the Majoron and active neutrino mass generation 
mechanisms. For this reason, we do not dub as ``minimal'' this realisation and in particular it violates the minimality 
condition iii) as two independent explicit symmetry-breaking terms are present, one for the LN and the other for the PQ.

The choice of the Lagrangian in Eqs.~\eqref{eq:TypeILNC} and \eqref{eq:TypeILNV} is clearly not the only possible UV completion. In particular, changing the PQ charge assignment in Eq.~\eqref{PQChargeAssignmentTypeILNCV} implies that $\phi^{(\ast)}$ insertions may be different. The spurionic approach adopted in Sect.~\ref{sec:Lagrangian} turns out to be very useful also in this case, to discuss the (formal) invariance of the different terms under the PQ symmetry and identify the minimal model. Thus, in what follows, we consider the general Lagrangian in Eq.~\eqref{eq:TypeILN} and apply the spurionic analysis for the PQ symmetry as we did for the LN in the previous section. 

First of all, given the structure of Eq.~\eqref{eq:TypeILN}, we notice that switching $N_R$ with $S_R$ gives physically equivalent configurations. Therefore, without any loss of generality, we simplify the spurionic discussion by fixing $PQ(L_L)=PQ(N_R)$, which implies the PQ invariance of the term proportional to $Y_N$. The spurionic charges of the other quantities entering Eq.~\eqref{eq:TypeILN} read as:
\be
\begin{aligned}
PQ(Y_S)=&PQ(N_R)-PQ(S_R)\,,\qquad 
&PQ(\Lambda_{NS})=&-PQ(N_R)-PQ(S_R)\\
PQ(\Lambda_{NN})=&-2PQ(N_R)\,,\qquad\quad
&PQ(\Lambda_{SS})=&-2PQ(S_R)\,.
\end{aligned}
\ee
In particular, if $PQ(\Lambda_{ij})=\pm PQ(\phi)$ then the corresponding term is a Yukawa-like interaction between the two HNL fields and the scalar $\phi^{(\ast)}$ (alike the terms proportional to $Y_{NN}$, $Y_{SS}$ and $Y_{NS}$ in Eq.~\eqref{eq:TypeILNC}); on the other hand, if $PQ(\Lambda_{ij})=0$ then we deal with a direct Majorana mass term; in all the other cases, $PQ(\Lambda_{ij})\neq0,\,\pm PQ(\phi)$ implies that it is not possible to write down the corresponding term at the renormalisable level.

We can now proceed with considering different hypotheses. First of all, if $PQ(N_R)=PQ(S_R)$, then $PQ(Y_S)=0$ and therefore also the second Yukawa term proportional to $Y_S$ is invariant under PQ. On the other hand, $PQ(\Lambda_{NN})=PQ(\Lambda_{NS})=PQ(\Lambda_{NN})=-2 PQ(N_R)$ and, by selecting $PQ(N_R)=-PQ(\phi)/2$,
we end up with the Lagrangian in Eq.~\eqref{eq:TypeILNC}, that is with a non-minimal model where the Majoron mass and the active neutrino masses are independent. Fixing $PQ(N_R)=+PQ(\phi)/2$ leads to an equivalent setup as indeed the corresponding Lagrangian is the one in Eq.~\eqref{eq:TypeILNC} by interchanging $\phi^\ast$ with $\phi$. On the other hand, for any other choice of $PQ(N_R)$, the second line of Eq.~\eqref{eq:TypeILNC} is strictly forbidden.

We therefore continue our discussion assuming that $PQ(N_R)\neq PQ(S_R)$. In this case, independently from the exact values of the charges, the Yukawa term proportional to $Y_S$ is not invariant under PQ. However, as already previously discussed, this term is necessary in order to obtain realistic active neutrino masses and therefore it must be introduced as an explicit breaking: we thus adopt the same notation as in the LSS with $Y_S\rightarrow \epsilon Y_{S}$ referring to the PQ symmetry breaking.

Next, if one of the two RH neutrinos has a vanishing PQ charge, then the corresponding Majorana HNL bilinear would be invariant under PQ and the associated term would be a direct mass. On the other hand, it would always be possible to fix the non-vanishing PQ charge of the other HNL such that a Yukawa-like Majorana term gets allowed in the Lagrangian. Explicitly, if $PQ(N_R)=0$, the term proportional to $\Lambda_{NN}$ is invariant under PQ and it enters the Lagrangian without any $\phi^{(\ast)}$ insertion. Then, there are two possibilities to give mass to the second HNL: either $PQ(S_R)=\mp PQ(\phi)/2$ or $PQ(S_R)=\mp PQ(\phi)$, corresponding to promoting to a Yukawa-like interaction the term with $\Lambda_{SS}$ or that with $\Lambda_{NS}$. At the Lagrangian level and using the notation of Eq.~\eqref{GenericNeutralMassLag}, these two cases read as
\be
\begin{aligned}
PQ(N_R)=0\quad\&\quad PQ(S_R)=\mp \dfrac{PQ(\phi)}{2}:\qquad
&\widehat{\Lambda}\longrightarrow
\begin{pmatrix}
\Lambda_{NN} & 0\\
0 & Y_{SS}\phi^{(\ast)}
\end{pmatrix}\\
PQ(N_R)=0\quad\&\quad PQ(S_R)=\mp PQ(\phi):\qquad
&\widehat{\Lambda}\longrightarrow
\begin{pmatrix}
\Lambda_{NN} & Y_{NS}\phi^{(\ast)}\\
Y_{NS}\phi^{(\ast)} & 0
\end{pmatrix}\,.
\end{aligned}
\ee
The opposite situation with $PQ(S_R)=0$ is very similar and would lead to 
\be
\begin{aligned}
PQ(S_R)=0\quad\&\quad PQ(N_R)=\mp \dfrac{PQ(\phi)}{2}\,:\qquad
&\widehat{\Lambda}\longrightarrow
\begin{pmatrix}
Y_{NN}\phi^{(\ast)} & 0\\
0 & \Lambda_{SS}
\end{pmatrix}\\
PQ(S_R)=0\quad\&\quad PQ(N_R)=\mp PQ(\phi)\,:\qquad
&\widehat{\Lambda}\longrightarrow
\begin{pmatrix}
0 & Y_{NS}\phi^{(\ast)}\\
Y_{NS}\phi^{(\ast)} & \Lambda_{SS}
\end{pmatrix}\,.
\end{aligned}
\ee
All these models, however, are not minimal as they violate the minimality condition ii) as there are two different scales associated with the Majorana terms after the PQ SSB: the direct Majorana mass and $f_a$.

On the other hand, if none of the HNLs has a vanishing PQ charge but $PQ(N_R)=-PQ(S_R)$, the diagonal Majorana terms would have charges $PQ(\Lambda_{NN})=-PQ(\Lambda_{SS})=-2 PQ(N_R)$ such that, taking $PQ(N_R)=\mp PQ(\phi)/2$, they can be written in the Lagrangian by multiplying by $\phi$ or $\phi^\ast$. On the other hand, $PQ(\Lambda_{NS})=0$ and it enters as a direct Majorana mass. At the Lagrangian level, we can write
\be
PQ(N_R)=-PQ(S_R)=-\dfrac{PQ(\phi)}{2}\,:\qquad
\widehat{\Lambda}\longrightarrow
\begin{pmatrix}
Y_{NN}\phi & \Lambda_{NS}\\
\Lambda_{NS} & Y_{SS}\phi^\ast
\end{pmatrix}
\ee
and equivalently for $PQ(N_R)=-PQ(S_R)=+PQ(\phi)/2$ interchanging $\phi^\ast$ with $\phi$ and viceversa. As for the previous two cases, also in this construction, there are two Majorana scales, $\Lambda_{NS}$ and $f_a$, and therefore the model is not minimal for condition ii). 

As the result of the discussion above, we further restrict the choice of the PQ charges of the HNLs such that $PQ(N_R)\neq \pm PQ(S_R)$ with both non-vanishing, preventing in this way any direct Majorana mass term in the Lagrangian. There are only two other possible setups that allow to give masses to both the HNLs and lead to the Seesaw mechanism. The first of them corresponds to promote the Majorana terms proportional to $\Lambda_{NS}$ and to $\Lambda_{SS}$ to be Yukawa-like interactions: the corresponding PQ charges and the Majorana block of the mass Lagrangian are
\be
PQ(N_R)=-\dfrac{PQ(S_R)}{3}=-\dfrac{PQ(\phi)}{2}\,:\qquad
\widehat{\Lambda}\longrightarrow
\begin{pmatrix}
0& Y_{NS}\phi^\ast\\
Y_{NS}\phi^\ast & Y_{SS}\phi
\end{pmatrix}
\ee
or the equivalent setup
\be
PQ(N_R)=-\dfrac{PQ(S_R)}{3}=+\dfrac{PQ(\phi)}{2}\,:\qquad
\widehat{\Lambda}\longrightarrow
\begin{pmatrix}
0& Y_{NS}\phi\\
Y_{NS}\phi & Y_{SS}\phi^\ast
\end{pmatrix}\,.
\ee
This construction does not suffer from the presence of multiple Majorana scales, as after the PQ SSB the non-vanishing entries are proportional to $f_a$, and thus it looks promising. However, the active neutrino mass matrix receives two different contributions at the tree level,
\be
m_\nu \simeq - \sqrt2\dfrac{Y_{SS}}{Y^2_{NS}}\dfrac{m_N\,m_N^T}{f_a} - \sqrt2\epsilon\dfrac{m_N \,m_S^T + m_S \, m_N^T}{Y_{NS} f_a} \,,
\label{ActiveNuMassLSSIV}
\ee
where the first term dominates, unless specific tuning is present among the parameters. It follows that it is not possible to correctly describe the active neutrino masses and PMNS mixing as the dominant term has rank 1, thus ruling out this model. 

\subsection{The minimal massive Majoron Seesaw Lagrangian} 
\label{sec:mmMlag}
The only case left unexplored identifies the minimal massive Majoron Seesaw (mmM) model where the Majoron mass and the active neutrino masses are indeed correlated. This model satisfies the three minimality conditions i)--iii) and describes realistic active neutrino masses and PMNS mixing. The PQ charges of the fields involved satisfy to 
\be
PQ(L_L) = PQ(N_R) = - \dfrac{PQ(S_R)}{3}=-\dfrac{PQ(\phi)}{2}
\label{PQChargesofmmMmodel}
\ee
and the corresponding PQ conserving and explicitly violating Lagrangian densities read as
\begin{align}
\label{eq:LagPQ}
-\sL^\text{mmM}_\text{PQ}= & \ov{L_L}\,\tH\, Y_N\,  N_R + \frac{Y_{NS}}{2}\,\phi^\ast\, \left(\ov{N_R^{c}}\, S_R + \ov{S_R^{c}}\,N_R\right) +
\dfrac{Y_{NN}}{2}\, \phi\,\ov{N_R^c}\,N_R + \hc \, , \\
-\sL^\text{mmM}_{\epsilon \text{PQ}}= & \epsilon\ov{L_L}\,\tH\, Y_S\,S_R  + \hc \,,     
\label{eq:LagPQV}
\end{align}
where as usual $Y_N, Y_{NN}$ and $Y_{NS}$ are assumed to be order one, while $\epsilon Y_S$ much smaller. Notice that this Lagrangian is invariant under the interchange of $\phi$ and $\phi^\ast$ as far as the sign of $PQ(\phi)$ in Eq.~\eqref{PQChargesofmmMmodel} is accordingly flipped.

Focussing on the explicit PQ breaking, it is straightforward to check that neglecting the $\epsilon Y_S$ term, the Majoron dependence can be removed from these Yukawa-like interactions, reappearing only in derivative couplings, ending again with a massless Majoron model. Moreover, in this same limit, the active neutrino mass matrix has rank 1 and cannot generate the two observed neutrino mass differences. On the other hand, once this term is taken into consideration, it explicitly breaks both the PQ symmetry and the LN: the Majoron acquires a mass and the active neutrino masses can be described according to the observations, both types of masses being necessarily proportional to $\epsilon Y_S$. 

These equations closely look like the expressions in Eqs.~\eqref{eq:neutrinosector} and \eqref{eq:neutrinosectorV} of the Extended Seesaw context, and indeed, after the SSB of the EW and PQ symmetries, the resulting lepton mass matrix matches the one in Eq.~\eqref{MajoronNeutralMassMatrix}, with the mass terms explicitly given by 
\begin{equation}
m_{N,S}= \frac{Y_{N,S}}{\sqrt{2}} v_\text{EW}\, , \qquad \qquad \Lambda_{NN,NS}= \frac{Y_{NN,NS}}{\sqrt{2}} f_a\, , \qquad \qquad
\Lambda_{SS}=0\,.
\label{DEFmM2}
\end{equation}

We can now use the experimental data from neutrino oscillation experiments, adopting for definiteness the results presented in Ref.~\cite{Esteban:2020cvm} (including the SK atmospheric data) to constrain the parameter space of the mmM model by use of Eqs.~\eqref{eq:mplusmminus}-\eqref{eq:IOdef}.
\begin{figure}[t!]
\centering
\includegraphics[width=1.0\textwidth]{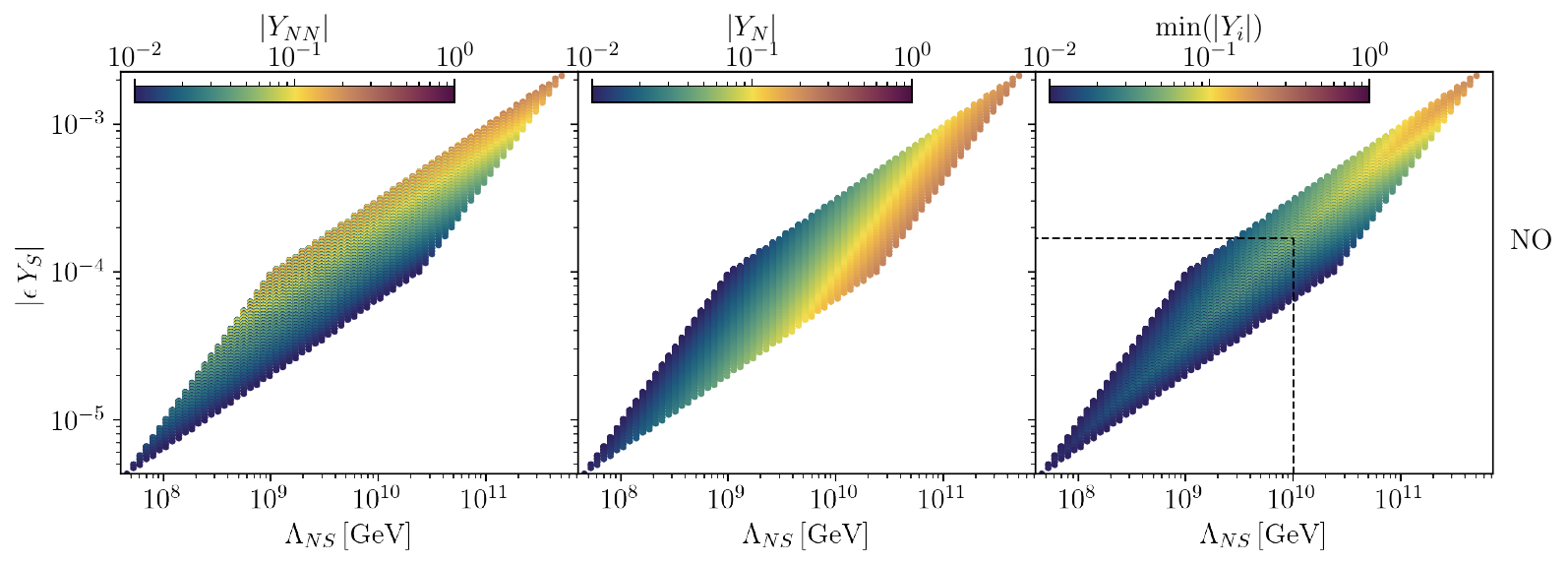}
\includegraphics[width=1.0\textwidth]{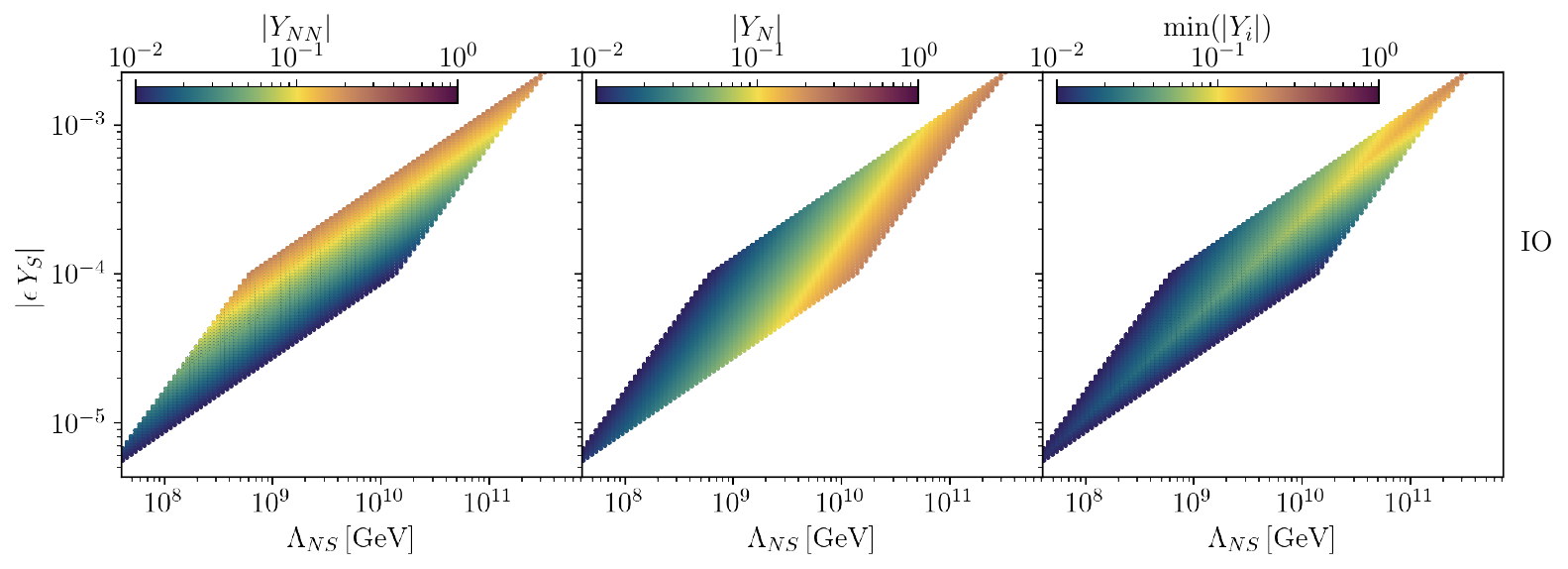}
\caption{\small 
The colored region in the plots represents the allowed $\{\epsilon ,\,Y_i,\,\Lambda_{NS}\}$ parameter space 
for which the condition $m_L \le 0.1 \,m_T$ is satisfied, assuming Normal Order (\textit{top}) or Inverse Order (\textit{bottom}), respectively. 
In the third upper plot the benchmark point $|\epsilon Y_S|=1.7\times 10^{-4},\,|Y_{NN}| = |Y_N| = 0.06$, and 
$\Lambda = 10^{10}~\mbox{\rm GeV}$ is shown.}
\label{fig:Parameter_Scan}
\end{figure}
In Fig.~\ref{fig:Parameter_Scan} we scan the parameter space $\Lambda_{NS}$ {\it vs.} $\epsilon Y_S$, running over the different Yukawa couplings taken in the ``natural'' range $|Y_i| \in \left[10^{-2} ,\,  1\right]$, letting $\epsilon$ the only \textit{ad hoc} ``small'' parameter. Their magnitude is represented by the coloured horizontal band: the smallest values are in blue, while the largest ones are in orange.  In the coloured regions, the mass eigenstates are fixed to reproduce both mass differences, by setting $r$ and one mass splitting to their experimental central values, while variations within the corresponding $3\sigma$ ranges do not show qualitative changes. The loop contribution is required to be at most the $10\%$ of the tree-level ones, to preserve the predictivity required by the hypothesis of this work. Moreover, $\epsilon$ satisfies the condition $\epsilon \leq 10^{-2} \min{|Y_i|}$, reflecting the soft explicit breaking of the LN. 

These conservative conditions show a parameter space where the scale $\Lambda_{NS}$ spans a relatively small range of values, $\Lambda_{NS} \sim 
10^8 - 10^{11}\GeV$, where the Yukawa couplings are larger than $10^{-2}$. As expected, relaxing any of the previous conditions enlarges the parameter space: e.g., requiring the loop contributions to be at most the $30\%$ of the tree-level one, one would allow reaching scales of $\Lambda_{NS} \sim10^{12}\GeV$ 
with Yukawa couplings of order $|Y_i|\sim 0.7$. 

Notice that the requirement of not going below $10^{-2}$ in the Yukawa couplings constrains the region of smaller $\Lambda_{NS}$ values. This can be seen in the left and centre plots of Fig.~\ref{fig:Parameter_Scan}, where $|Y_N|$ and $|Y_{NN}|$ need to be small in order 
to tame the loop contribution. As $\Lambda_{NS}$ grows the loop contribution stops being so relevant, but as $\epsilon$ necessarily grows 
to fix the correct mass splitting, the constraint of $\epsilon \leq 10^{-2}\min{|Y_i|}$ exclude the upper sides of the rhomboid.

The next two sections are devoted to the study of the Majoron mass and its phenomenological constraints.

%
%
\section{One-loop contributions to the Majoron mass}
\label{sec:MajoronMass}

Before calculating the one-loop contributions to the Majoron mass in the mmM scenario, it is useful to discuss a simpler model, similar 
to the one introduced in \cite{Hill:1988bu}, to highlight some fundamental features. 

\subsection{Pseudo-GB mass radiative contributions in a toy model}
\label{sec:toymodel}

We consider a Dirac fermion field $\psi$ coupled to a complex scalar singlet $\phi$ through the interaction Lagrangian 
\begin{align}
	-\sL = y\, \phi\,\ov{\psi}_L\,\psi_R + \tilde{y}\,\phi^\ast\,\ov{\psi_L}\psi_R+\hc
	\label{eq:ToyLagrangian}
\end{align}
with $y$ and $\tilde{y}$ reals. The fields $\psi$ and $\phi$ have non-trivial transformation properties under a global $U(1)$, spontaneously broken by $\phi$ getting a non-vanishing vev, $f_a$. It is clear from the above Lagrangian that, due to the simultaneous presence of $y$ and $\tilde{y}$, there is no possible charge assignment for which the interaction Lagrangian could preserve any $U(1)$ symmetry. One could assume that $y$-term is the symmetry preserving coupling and $\tilde{y}$-term is the (small) softly breaking one, but the opposite assumption is viable and leads to the same physical results. For the time being, without fixing a specific charge assignment and thus without identifying which term among $y$ and $\tilde y$ is symmetry breaking, we just consider their product $y\,\tilde y$ to be a small quantity. 

After SSB has occurred, the fermion mass term and coupling with the pseudo-GB $a$ reads:
\begin{align}
-\sL 
&\supset m_\psi\cos\left(\dfrac{a}{f_a}\right)\ov{\psi}\,\psi + i\, m'_\psi\sin\left(\dfrac{a}{f_a}\right)\ov{\psi}\,\gamma_5\,\psi\,, \\
&\supset m_\psi\,\ov \psi\psi + i\,m'_\psi\dfrac{a}{f_a} \ov\psi\,\gamma^5\psi-\dfrac{m_\psi}{2}\left(\frac{a}{f_a}\right)^2\ov\psi\psi 
+ \dots \,, 
\label{eq:massesToyLag}
\end{align}
where $m_\psi$ and $m'_\psi$ are defined as
\be
m_\psi\equiv \dfrac{y+\tilde{y}}{\sqrt2}\, f_a\,,\qquad\qquad  m'_\psi\equiv\dfrac{y-\tilde{y}}{\sqrt2}\, f_a \,.
\ee
and in the second line, we have expanded in $a/f_a$ and kept the two lowest order contributions that are the only relevant for a 
one-loop calculation. 

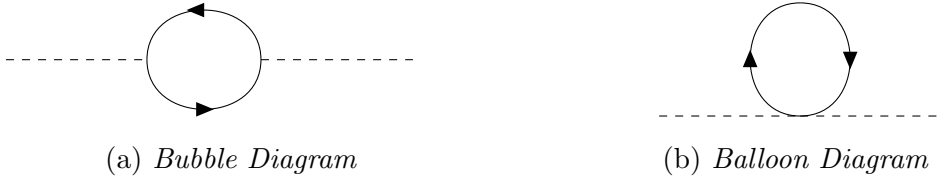
\begin{figure}[t!]
\centering
\begin{subfigure}[b]{0.4\textwidth}
\begin{tikzpicture} 
\begin{feynman}
\centering
\vertex(a){};
\vertex[right=2cm of a](b);
\vertex[right=1.5cm of b](c);
\vertex[right=2cm of c](d);
\diagram* {
(a) -- [dashed] (b),
(b) -- [anti fermion, half left] (c),
(b) -- [fermion,half right](c),
(c) -- [dashed] (d),
};
\end{feynman}
\end{tikzpicture}
\caption{\em Bubble Diagram}
\end{subfigure}
\begin{subfigure}[b]{0.5\textwidth}
\centering
\begin{tikzpicture}
\begin{feynman}
\centering
\vertex(a){};
\vertex[right=2cm of a](b);
\vertex[right=2cm of b](c);
\vertex[above=1.5cm of b](d);
\diagram* {
(a) -- [dashed] (b),
(b) -- [fermion, half left] (d),
(d) -- [fermion, half left] (b),
(b) -- [dashed] (c),
};
\end{feynman} 
\end{tikzpicture}
\caption{\em Balloon Diagram}
\end{subfigure}
\caption{\small Diagrams contributing to the 1-loop mass of the GB.}
\label{fig:GBdiagrams}
\end{figure}

As the simultaneous presence of $y$ and $\tilde{y}$ explicitly breaks the $U(1)$ symmetry, one expects a GB mass term to appear at 
loop level proportional to the product $y\,\tilde{y}$. Denoting with $-i\sM(p^2)$ the sum of all one-particle-irreducible one-loop contribution to the GB scalar propagator, any 
positive contribution to $\sM(p^2=0)$ corresponds to a positive shift of the GB mass. Such contributions can appear only from two 
types of loops, shown in Fig.~\ref{fig:GBdiagrams}: a) the Bubble diagram and b) the Balloon diagram. The Bubble diagram (\textit{left}) needs 
two 3-point vertices, and therefore is proportional to $m^{\prime 2}_\psi$, while the Balloon one (\textit{right}) only involves a single 
4-point vertex and is proportional to $m_\psi$. 

In dimensional regularization (with $d\equiv4-2\varepsilon$), one obtains that the zero momentum contribution to the GB propagator reads:
\be
\begin{aligned}
-i\sM=&-i\sM_A-i\sM_B\\
=& \,i\,\dfrac{m_\psi^{\prime2}}{f_a^2} \int \dfrac{\dd^dk}{(2\pi)^d}\dfrac{\Tr{\left[\gamma^5 (\slashed{k}+m_\psi)\gamma^5 
(\slashed{k}+m_\psi)\right]}}{(k^2-m_\psi^2)^2}+i\,\dfrac{m_\psi}{f_a^2} \int \dfrac{\dd^d k}{(2\pi)^d}\dfrac{\Tr{\left[\slashed{k}+m_\psi\right]}}{k^2-m_\psi^2} \\ 
=& -i\,\dfrac{(-m_\psi^2\,m_\psi^{\prime2}+m_\psi^4)}{4\pi^2 f_a^2}\left[1-\log\left(\dfrac{m_\psi^2}{\mu_R^2}\right)+\dfrac{1}{\tilde\varepsilon_{UV}}\right]
\end{aligned}
\label{toymodelResultsChiFlipping}
\ee
with the first (second) term referring to the bubble (balloon) diagram. In Eq.~\eqref{toymodelResultsChiFlipping}, $\mu_R$ is the renormalisation 
scale and $\tilde{\varepsilon}_{UV}$ is defined as
\be
\dfrac{1}{\tilde{\varepsilon}_{UV}} \equiv \dfrac{1}{\varepsilon_{UV}} - \gamma_E+\log(4\pi)\,,
\ee
with $\gamma_E$ the Euler-Mascheroni constant\footnote{Notice that the momentum zero computation is consistent with the fact that the GB is massless at tree level and therefore any correction to the kinetic term would effectively be translated into a 2-loop correction to its mass.}. The one-loop pseudo-GB mass, in the $\ov{\text{MS}}$-scheme, then reads
\be
m_a^2 = \dfrac{y \tilde{y} \left(y+\tilde{y}\right)^2 }{4\pi^2}f_a^2\,,
\label{eq:matoymodel}
\ee
confirming the dependence on the product between $y$ and $\tilde{y}$, independently from the $U(1)$ charge assignment, that is without having identified which one among $y$ and $\tilde{y}$ does explicitly break the symmetry. On the other hand, if any of the two parameters is 
vanishing the Lagrangian is left with an accidental symmetry that protects the GB from acquiring a mass and indeed the expression in Eq.~\eqref{eq:matoymodel} does vanish. This can be easily understood by writing the scalar field in the polar coordinates. Indeed, the two terms in Eq.~\eqref{eq:ToyLagrangian} would have exponentials with opposite signs and, only if either $y$ or $\tilde{y}$ is vanishing, then it is possible to perform a $\psi$ redefinition to reabsorb the GB dependence. In this case, the GB would reappear in the kinetic terms and only with derivative couplings, implying the presence of a shift symmetry that protects the GB from acquiring a mass term. 

It is instructive to repeat the computation of the GB mass in the chirality-preserving basis, also known in the literature as 
\textit{derivative basis}. Moreover, to make more evident the results of this exercise, we assume a specific charge assignment such that the $y$-term is $U(1)$-preserving, while $\tilde y$ is $U(1)$-breaking and we move to a more explicit notation with $\tilde{y} \to \epsilon \tilde{y}$, being $\epsilon$ a small parameter\footnote{Obviously the alternative assignment in which the symmetry is broken by the $y$-term would be perfectly equivalent.}. By performing the following field-dependent redefinition in the whole Lagrangian,
\be
\psi \rightarrow e^{-i\gamma_5 a/2f_a}\,\psi\,.
\ee
the GB dependence is removed from the $y$-term and is only left in the $\tilde y$-term, and the relevant interactions read
\be
\begin{aligned}
	-\sL&\supset \dfrac{y\,f_a}{\sqrt2}\ov{\psi}\psi + \epsilon\dfrac{\tilde{y}\,f_a}{\sqrt2}\left[\cos\left(\dfrac{2a}{f_a}\right)\ov{\psi}\psi- 
	i\,\sin\left(\dfrac{2a}{f_a}\right)\ov{\psi}\gamma_5\psi\right] -\dfrac{\partial_\mu a}{2f_a}\,\ov{\psi}\gamma^\mu\gamma_5\psi + \cdots \\
	&\supset \dfrac{(y+\epsilon\tilde{y})\,f_a}{\sqrt2}\,\ov{\psi}\psi-\epsilon\dfrac{2\,\tilde{y}\,f_a}{\sqrt2}\left[\left(\frac{a}{f_a}\right)^2\ov{\psi}\psi+
	i\,\dfrac{a}{f_a}\ov{\psi}\gamma_5\psi\right] -
	\dfrac{\partial_\mu a}{2f_a}\,\ov{\psi}\gamma^\mu\gamma_5\psi + \cdots \,, 
\end{aligned}
\label{toy:derivative-basis} 
\ee
where in the last line only the linear and quadratic terms in $a/f_a$ have been kept. For a vanishing $\tilde{y}$, the $a$ field has only derivative couplings reflecting its exact GB nature. Instead, for $\tilde{y}$ small, but different from zero, shift-breaking terms are present and lead to the one-loop pseudo-GB mass term. 

In the derivative basis, the computation of the mass is much simpler as indeed i) diagrams with (one or two) derivative couplings never 
contribute to the GB mass as they always carry an external momentum  dependence, and ii) the GB mass contribution is dominated by the balloon diagram, 
\be
-i\sM_B
=-i\,\epsilon\,\dfrac{y^3\, \tilde{y}\,f_a^2}{4\pi^2}\left[1-\log\left(\dfrac{m_\psi^2}{\mu_R^2}\right)+\dfrac{1}{\tilde\varepsilon_{UV}}\right]+
\cO\left(\epsilon^2\right)\,,
\label{toymodelResultsChiPreserving}
\ee
being the bubble one suppressed by two powers of the small parameter $\epsilon$. By expanding Eq.~\eqref{toymodelResultsChiFlipping} 
in $\epsilon$ and keeping the leading term, $(-m_\psi^2\,m_\psi^{\prime2}+m_\psi^4)=y^3\tilde{y} f_a^4$ and we recover the result in Eq.~\eqref{toymodelResultsChiPreserving}. This proves that moving to the derivative/chirality preserving basis, leaving the GB dependence only into the explicitly breaking term(s), is the most convenient one for what concerns the calculation of the contributions to the GB mass. Indeed, the computations reduce to only one topology of diagrams, the Balloon one, at the leading order in $\epsilon$. This is indeed the choice that we will make in the next section for the mmM model.

\subsection{Majoron mass in the Minimal Majoron model}
\label{sec:radiative}

In this section we calculate the one-loop contribution to the Majoron mass in the mmM model and the complete Yukawa Lagrangian relevant for this computation is the following:
\be
\begin{split}
-\sL_{\text{Yuk}} = & \phantom{+}\ov{L_L}\,H \, Y_e\, e_R + \ov{L_L}\,\tH\, Y_N\,  N_R + \epsilon \,\ov{L_L}\,\tH\,Y_S\,S_R + \\
& + \dfrac{Y_{NS}}{2}\, \phi^\ast\, \left(\ov{N_R^{c}}\, S_R+\ov{S_R^{c}}\, N_R \right)+ \dfrac{Y_{NN}}{2}\, \phi\,\ov{N_R^c}\,N_R + \hc
\end{split}
\label{eq:neutrinosectortot}
\ee
Once LN is spontaneously broken, the $\phi$-radial model, $\rho$, acquires a large mass $m_\rho \approx f_a\gg v_\text{EW}$, and therefore it can be safely integrated out and it is not expected to have any significant impact in the low-energy phenomenology we are interested in. In the following, only the lepton couplings with the light angular mode, $a$, are considered.

Armed with the toy-model discussion, the simplest approach to calculate the Majoron mass contributions is moving to the chirality preserving basis, leaving the Majoron dependence on the explicit  PQ breaking term, where only the balloon diagram contributes at leading order in the small parameter $\epsilon$. By means 
of the following field-dependent redefinition of the fermionic fields,
\be
\{N_R,\,L_L,\,e_R\}\rightarrow \{N_R,\,L_L,\,e_R\}\,e^{-ia/(2f_a)}\,,\qquad\qquad S_R\rightarrow S_R\,e^{3ia/(2f_a)}\,,
\label{eq:fieldredefinition}
\ee
the Majoron dependence in the second-line terms in Eq.~\eqref{eq:neutrinosectortot} is reabsorbed, without reappearing in the first two terms (i.e.~the PQ conserving ones) of the first line. The only dependence on the Majoron field, after the redefinition 
in Eq.~\eqref{eq:fieldredefinition}, is left in the only PQ symmetry violating term of the first line, i.e. the one proportional to $\epsilon Y_S$, and in the derivative couplings that originate from the fermion kinetic terms: all in all, the Lagrangian containing the Majoron interactions reads
\be
	\sL_{a}= \dfrac{\derp_\mu a}{2f_a}\left( \ov{\nu}_L\gamma^\mu \nu_L + \ov{N}_R\gamma^\mu N_R 
	-3\,\ov{S}_R\gamma^\mu S_R\right)\, + \left(\epsilon\, \ov{\ell}_L\,\tH\,Y_S\,S_R\,e^{2ia/f_a}+\hc\right) \,.
\label{MajoronLagChiPreserving}
\ee
As all the SM leptons identically transform under the PQ, there is no tree-level Majoron coupling with the charged leptons and no anomalous gauge terms.

Before proceeding with the calculation of the one-loop contribution to the Majoron mass, it is convenient to write the Lagrangian in terms 
of the physical fields, that is the mass eigenstates of the neutral lepton mass matrix Eq.~\eqref{MajoronNeutralMassMatrix} accounting for the parameter definitions in Eq.~\eqref{DEFmM2}. All the details of the straightforward procedure are reported in App.~\ref{app:MassEigen} and for convenience, we only report here the leading contributions to the HNL masses in terms of the fundamental parameters of the mmM Lagrangian, 
\be
M_{N,S}=\dfrac{\Lambda_{NS}}{2}\left[\sqrt{4+\left(\dfrac{Y_{NN}}{Y_{NS}}\right)^2}\mp 
\left(\dfrac{Y_{NN}}{Y_{NS}}\right)\right]\,.
\ee

As in the chirality preserving basis the leading order contribution to the Majoron mass comes from the Balloon type diagram, the leading contribution to the Majoron mass comes from
\be
-\sL_{a} \supset  \,\dfrac{|m_N|| \epsilon m_S||\eta|}{2\sqrt{M_N\,M_S}
	(M_N+M_S)}\Bigg(M_N\ov{S_R^c}S_R-M_S\ov{N_R^c}N_R\Bigg)\,\frac{a^2}{f_a^2} +\hc\,.
\label{FinalChiPreservingBasisRelevantText} 
\ee
whose derivation can be found in App.~\ref{app:MassEigen}. In fact, the SM neutrino contributions to the Majoron mass are proportional to the active neutrino masses and therefore are completely negligible. Moreover, the $\ov{S_R^c}N_R$ terms come with a linear coupling to the 
Majoron, and therefore, as learned from the toy model analysis, they contribute to $m_a$ only through the bubble type of diagrams, 
thus being of $\mathcal{O}(\epsilon^2)$. However, in this model, a stronger statement regarding the Majoron mass can be made. The potential is constrained by symmetry arguments to be of the form
\begin{equation}
\label{eq:general-scaling-potential}
    V(a) \propto |m_N| |\epsilon m_S| \Lambda_{NN}\Lambda_{NS}\,\cos\left(\dfrac{2a}{f_a}\right)\,,
\end{equation}
meaning that dimensionally the potential is already saturated by the necessary combination of EW- and Majorana-Yukawas (cfr.~Eq.~\eqref{TrM4logM2}).
The presence of extra EW-Yukawa factors in general, including dependence on $\epsilon$, can only enter as a correction factor of the type $\propto (1+|m_{N,S}|^2/f_a^2)$, meaning that it would be NLO in $1/f_a$ expansion. This implies that in the derivative basis, the balloon captures the full result of the amplitude at LO in $f_a$. The same statement is not valid in the toy model, where only one scale is present and thus no suppressions of the type $v_\text{EW}/f_a$ are possible.

From the couplings in Eq.~\eqref{FinalChiPreservingBasisRelevantText}, one can calculate the Balloon contribution to the Majoron 
mass in the $\overline{\text{MS}}$ scheme. The result can also be derived employing the CW potential, as shown in App.~\ref{app:CW} in Eq.~\eqref{MajoronMassCW}. Assuming $v_{EW} \ll f_a$, in the NO case, the Majoron mass reads
\begin{align}
m_a^2 =  \dfrac{ |\eta||m_N|| \epsilon m_S|}{\pi^2 } \dfrac{\sqrt{M_N M_S}}{M_N+M_S}
	&\Bigg[\frac{\left(M_S^2+M^2_N\right)}{f_a^2}\log\left(\dfrac{M_S}{M_N}\right)+\nn\\
	&\hspace{0.5cm}+ \frac{(M_S^2-M_N^2)}{f_a^2}\left(\log\left(\dfrac{M_N M_S}{\mu^2_R}\right)-1\right)\Bigg]\, .
 \label{eq:mass-before-constraint}
\end{align}
Let us notice that in a generic model with explicit PQ symmetry breaking one expects typically $m_a^2 \propto \epsilon f_a^2$. This 
happens, indeed, in the toy model described in Sec.~\ref{sec:toymodel} as clearly revealed by Eq.~\eqref{toymodelResultsChiPreserving}. 
Instead, with the specific symmetry-breaking pattern introduced for the mmM model, the Majoron mass for large $f_a$ behaves as 
$m_a^2 \propto \epsilon \, v_\text{EW}^2 \log{f_a/\mu_R}$, that is, $m_a^2$ asymptotically depends only logarithmically from the large 
PQ SSB scale, thus allowing a naturally lighter ALP\footnote{See also the discussion in Ref.~\cite{Frigerio:2011in}.}.

Assuming now also $m_L \sim m_{T_2} \ll m_{T_1}$ and by means of Eqs.~\eqref{eq:mplusmminus} and \eqref{eq:NOdef}, the Majoron mass 
can be strictly connected with the neutrino mass splittings
\begin{align}
 m_a^2\simeq\dfrac{\sqrt{|\eta||\Delta m_{32}^2|}}{2\,\pi^2 } \dfrac{M_N M_S}{M_N+M_S}
	&\Bigg[\frac{\left(M_S^2+M^2_N\right)}{f_a^2}\log\left(\dfrac{M_S}{M_N}\right)+\nn\\
	&\hspace{0.5cm}+ \frac{(M_S^2-M_N^2)}{f_a^2}\left(\log\left(\dfrac{M_N M_S}{\mu^2_R}\right)-1\right)\Bigg]\,,
\label{eq:MajoronMassExplicit}
\end{align}
where, to understand the correct scale dependence in $m_a^2$ one has now to recall from Eq.~\eqref{eq:mT1}  that $\Delta m^2\propto 
v_\text{EW}^4/f_a^2$. The result for the IO scenario can be obtained by simply replacing $|\Delta m_{32}^2| \to |\Delta m_{21}^2|$ 
in Eq.~\eqref{eq:MajoronMassExplicit}. 
\begin{figure}[t]
\centering
\includegraphics[width=\textwidth]{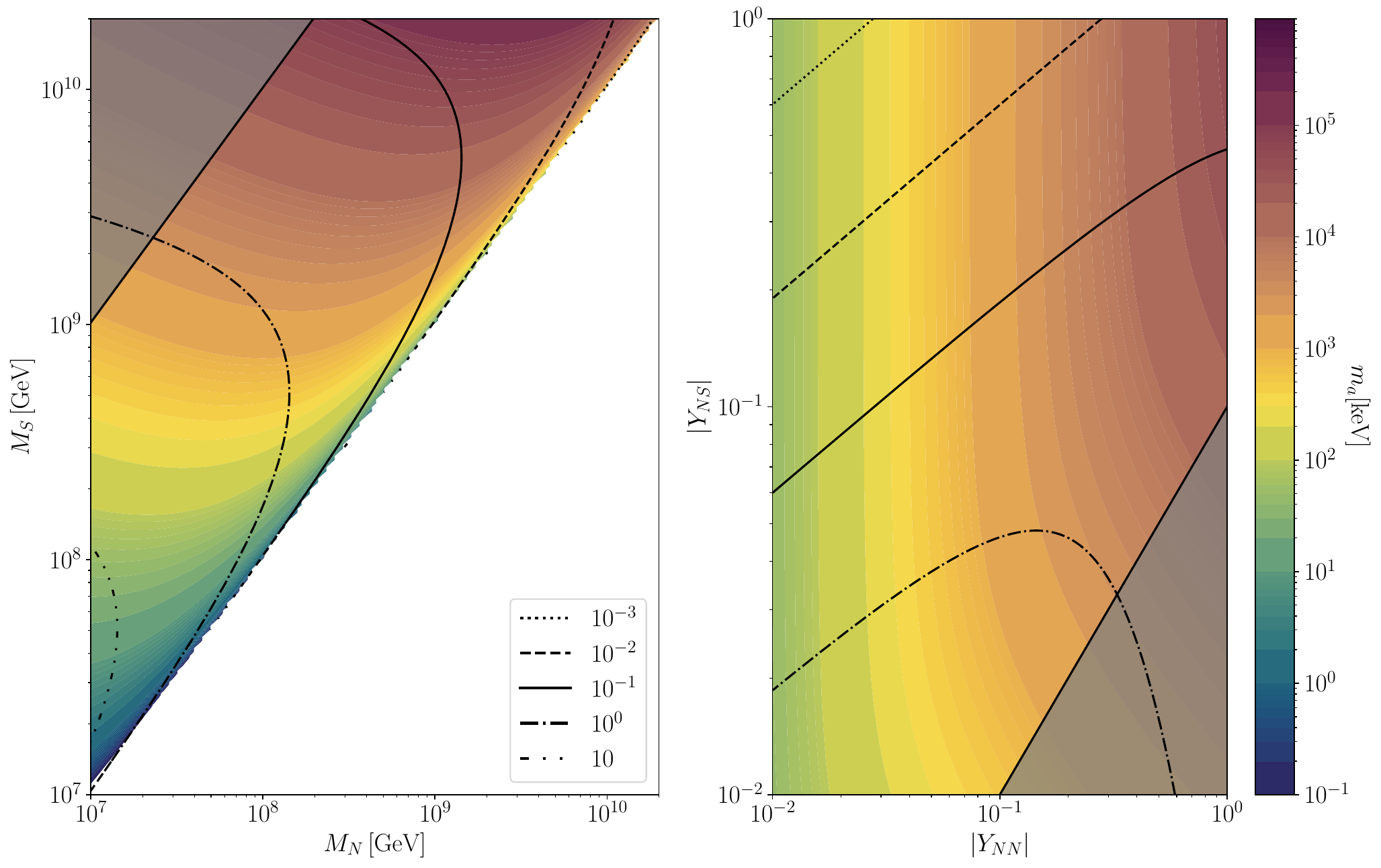}
\caption{\small
Dependence of the Majoron mass with respect to the (\textit{left}) HNL masses $M_{N,S}$ and (\textit{right}) the 
Yukawas $Y_{NN,NS}$, for $f_a=10^{10}\,$GeV and $Y_N=0.01$. The different colours indicate the mass range of the Majoron, while the different black lines are contours of the  $m_L/\sqrt{\Delta m_{\text{atm}}^2}$ ratio. The gray area bounds the region where $m_{T_2}>0.1\,  m_{T_1}$.}
\label{fig:ContourMass}
\end{figure}
To provide an intuitive idea of the Majoron mass behaviour in terms of the relevant model parameters, in Fig.~\ref{fig:ContourMass} 
$m_a$ as a function of the HNLs masses $M_{N,S}$ (\textit{left}) and the Yukawas $Y_{NN,NS}$ (\textit{right}) has been shown, with the different 
colour nuances indicating the Majoron mass (in keV) as reported on the scale on the right of the figure. The different black 
contours indicate, instead, the corresponding $m_L/\sqrt{\Delta m_{ \text{atm}}^2}$ values as an estimation of the relative size of the 
loop contribution with respect to the tree-level one. This gives an idea, similarly to Fig.~\ref{fig:Parameter_Scan}, of how 
much the parameter space gets constrained for taming the one-loop contribution to the active neutrino masses. In grey, we estimate the region that could induce a $m_{T_2}>0.1 m_{T_1}$. In terms of the 
HNL masses (\textit{left plot}) this means that, in order to keep the one-loop vs tree-level ratio below $0.1$, one needs either going close to the mass degeneracy region or increasing $M_S$ above $\sim 10^9$ GeV for a fixed $M_N$, however, this last region would create a larger $m_{T_2}$ contribution. The white 
region in the left plot of Fig.~\ref{fig:ContourMass} corresponds to the inaccessible region once $M_S > M_N$ is chosen in 
Eq.~\eqref{eq:HNL-masses}. The right plot of Fig.~\ref{fig:ContourMass} represents the same information but as functions of 
the $Y_{NN,NS}$ Yukawas: one can notice here that for Yukawas $Y_{NN,NS} \sim \mathcal{O}(10^{-2})$ and for the chosen scale 
$f_a=10^{10}$ GeV, a Majoron mass around the keV scale is typically reproduced, while a $\sim 100\,$MeV scale Majoron can be obtained 
for $\mathcal{O}(1)$ Yukawas couplings.

\begin{figure}[t]
\centering
\includegraphics[width=15cm]{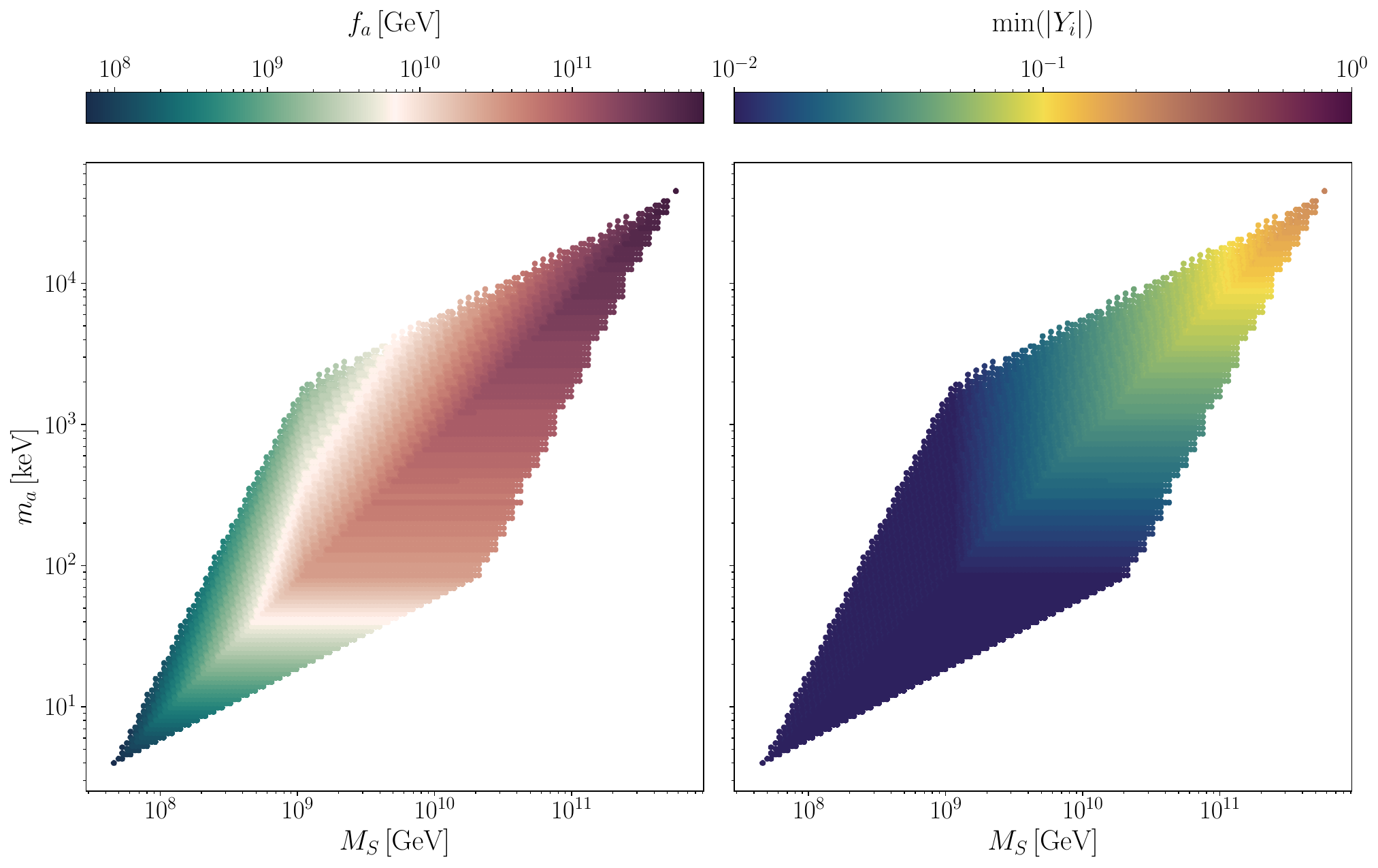}
\caption{\small
Majoron mass plotted as a function of $M_S$. Different colours indicate the dependence on the $f_a$ scale (\textit{left plot}), 
or on the $\min{|Y_i|}$ value (\textit{right plot}) as shown in the corresponding upper bars.}
\label{fig:mavsMS}
\end{figure}

The second task is to identify the parameter space that satisfies the constraints mentioned in Sect.~\ref{sec:mmMlag}, that is $\epsilon\ll 0.01\min{|Y_i|}$ and $m_{T_2}\, , m_{L} \ll m_{T_1}$. The allowed parameter space is then shown 
in Fig.~\ref{fig:mavsMS} with respect to one of the masses of the HNLs. The pair of HNLs needs to be non-degenerate, as otherwise $Y_{NN}=0$ and the Majoron would become massless. However, a small tuning needs to be employed to control the size of the loop level. For this reason, the pair of HNLs is quasi-degenerate and we can plot the dependence for only one of the masses. On the plots of Fig.~\ref{fig:mavsMS}
we observe the dependence on the scale $f_a$ (\textit{left}), which shows a linear scaling $m_a\sim f_a$, this dependence is clear only when one fixes $\epsilon |m_S| |m_N|$ to reproduce the neutrino masses, while if one keeps the dependence on the Yukawas (as in Eq.~\eqref{MajoronMassCW}) this scaling is implicit. The range of $f_a$ goes from $10^8-10^{11}\,$GeV for masses of the Majoron between $1-5\times10^4\,$keV. We also see in Fig.~\ref{fig:mavsMS} (\textit{right}) the dependence of the mass with respect to the Yukawas showing that the region of the largest Yukawas corresponds to the heaviest Majoron.  

Note that this constrained space can be enlarged at the cost of relaxing our ``natural'' Yukawas, however at the cost of needing to control both loop and tree level extra contributions. This would be possible to accommodate the neutrino masses, but a loss in the predictivity of the Majoron mass would be expected.

%
%
\section{Phenomenology of the minimal massive Majoron}
\label{sec:Pheno}
%
%
In the previous section, it has been shown that the typical Majoron mass range, predicted by the mmM model, lies in the
$\left[1,\, 5\times 10^4\right]$ keV interval, i.e. slightly below the muon mass, with the typical SSB scale ranging 
between $\left[10^{7}, 10^{11}\right]$ GeV. In this section, the main phenomenological impact of the mmM model in particle 
and astroparticle/cosmological observables will be discussed.

The full mmM Lagrangian in the neutral lepton mass basis is described in App.~\ref{app:Majoron-interactions}. At tree-level the Majoron 
couples exclusively with neutral leptons, and clearly only the Majoron couplings with active neutrinos can be directly constrained 
by present experiments. Imposing that the observed neutrino oscillation mass differences and PMNS mixing angles are being reproduced by 
opportunely choosing the Dirac and Majorana Yukawas entries of the neutral lepton mass matrix, the lowest order\footnote{Alternatively 
here one could use as $m_\nu$ the complete tree-level plus one-loop neutrino mass matrix calculated in Eqs.~\eqref{ActiveNuMass} 
and \eqref{eq:numassloop}.} Majoron-active neutrino coupling (see for example Eq.~\eqref{eq:Lagmassbasis}) reads
\be
\sL_{a\nu\nu}=-\dfrac{i\,a}{2f_a}\ov{\nu_L}\, m_\nu \,\gamma_5\,\nu_L^c\,.\\
\ee 

Couplings of the Majoron with charged SM leptons arise at one loop level through Z and W exchanges. For having a more compact 
notation, and matching with the existing literature~\cite{Garcia-Cely:2017oco,Heeck:2019guh}, it is useful to introduce the 
following adimensional hermitian coupling 
\begin{equation}
    K \equiv \dfrac{\hat{m}\hat{m}^\dagger}{v^2_{EW}}=\dfrac{m_N m_N^\dagger+\epsilon^2 m_S m_S^\dagger}{v^2_{EW}} = 
    \dfrac{1}{2}\left(|Y_N|^2 uu^\dagger+\epsilon^2|Y_S|^2 vv^\dagger\right)\,.
\label{eq:K}
\end{equation}
Since in our model the RH leptons couple simultaneously to $\phi$ and $\phi^*$, see Eq.~\eqref{eq:neutrinosectortot}, the Majoron-charged 
lepton couplings derived in \cite{Garcia-Cely:2017oco,Heeck:2019guh,Herrero-Brocal:2023czw} need to be accordingly modified. Therefore, 
the one-loop Majoron-charged leptons effective Lagrangian in the mmM model reads:
\be
\sL_{a\ell\ell}=\dfrac{i\,a}{16\pi^2 f_a}
\ov{\ell}\left(M_\ell \,\tr\left[\widetilde{K}_1 \right] \gamma_5 + 2 \,M_\ell\, \widetilde{K}_2\,P_L -2\, \widetilde{K}_2^\dagger\, M_\ell \, P_R\right)\ell\,,
\label{eq:MajoronLeptonLagrangian}
\ee
where $M_{\ell}$ is the diagonal charged leptons mass matrix and we have defined
\begin{align}
    &\widetilde{K}_1=K+\dfrac{(\sigma-1)}{v_\text{EW}^2}\left[m_N m_N^\dagger\left(1-\dfrac{2}{3}R^2\right)-R\,\epsilon m_Sm_N^\dagger+\dfrac{2}{3}R^3\,m_N\epsilon m_S^\dagger +\epsilon^2m_S m_S^\dagger\left(1+\dfrac{2}{3}R^2\right)\right]\,,\nn\\
    &\label{eq:Ktilde}\\[-3mm]
    &\widetilde{K}_2=K+\dfrac{(\sigma-1)}{v_\text{EW}^2}\left[m_N m_N^\dagger\left(1-\dfrac{5}{12}R^2\right)-R\left(\dfrac{5}{4}+\dfrac{1}{6}R^2\right)\,\epsilon m_Sm_N^\dagger\right.\nn\\
    &\qquad \qquad \qquad \qquad \qquad \qquad \qquad \left.+R\left(\dfrac{1}{4}+\dfrac{7}{12}R^2\right)\,m_N\epsilon m_S^\dagger +\epsilon^2m_S m_S^\dagger\left(1+\dfrac{5}{12}R^2\right)\right]\,.\nn
\end{align}
In Eq.~\eqref{eq:Ktilde} the 
parameter $\sigma = \pm 1$ has been introduced to switch easily between the models usually described in the literature 
($\sigma=1$) where the RH leptons are coupled solely to $\phi$, from our model ($\sigma=-1$) where they are coupled simultaneously to 
$\phi$ and $\phi^*$. Notice also that for $\sigma=-1$ but $R\ll 1$, one gets $\widetilde{K}_{1,2}\simeq-K$, that is, exactly the same 
coupling of the models with $\sigma=1$ but with $a\to-a$. In other words, the $R\ll 1$ region of our model corresponds to a scenario where the 
RH leptons couple only to $\phi^\ast$, as emerges looking at the Lagrangian of Eq.~\eqref{eq:neutrinosectortot}. Finally, to match 
with (part of) the existing literature, bounds to the Majoron-electrons coupling are going to be expressed in terms of the dimensional 
parameter, $g_{ae}$, defined as
\begin{equation}
    \label{eq:gaedefSP}
    \sL_{aee}= i\frac{m_e}{16\pi^2 f_a} \bar e\left[\Tr{\widetilde{K}_1}\gamma_ 5-2\left(i\Im{\widetilde{K}_{2,ee}}-\gamma_5 \Re{\widetilde{K}_{2,ee}}\right)\right] a \, e \equiv 
    \, \, a \,\bar e \,\left(g_{ae}^S+g^P_{ae}i\gamma_ 5\right) \,e\,.
\end{equation}
In general, the scalar coupling $g_S$ is proportional to
\begin{align}
    &\dfrac{\widetilde{K}_2-\widetilde{K}_2^\dagger}{2}=\dfrac{3}{4}\dfrac{(\sigma-1)}{v_\text{EW}^2}R\left(1+\dfrac{1}{2}R^2\right)\left[m_N\epsilon m_S^\dagger-\epsilon m_Sm_N^\dagger\right]\,,
\end{align}
which is $R$-suppressed with respect to the Hermitian part of $\widetilde{K}_{2}$, $\widetilde{K}_{2}^H\equiv(\widetilde{K}_2+\widetilde{K}_2^\dagger)/2$. As it turns out, in 
the allowed parameter space the exact one-loop calculation is fairly approximated by the $R\ll 1$ expression. This implies that for phenomenological purposes, only $g^P_{ae}$ is relevant, thus we can approximate
\begin{equation}
    \label{eq:gaedef}
    \sL_{aee}\approx 
    i\, \, a \,\bar e \,g^P_{ae}\gamma_ 5 \,e= i\frac{m_e}{16\pi^2 f_a} \left(\Tr{\widetilde{K}_1}-2(\widetilde{K}_2^H)_{ee}\right) a\, \bar e \,\gamma_ 5 \, e \equiv i g_{ae}\, \, a \,\bar e \gamma_ 5 \,e\,.
\end{equation}

We are not reporting here Majoron couplings with quarks and nucleons as the associated phenomenology, in our mmM scenario, is less 
compelling then the charged leptons one. In fact, Majoron couplings with quarks are flavour diagonal and therefore relevant bounds 
from flavour changing neutral currents observables (for example in $s\to d \,a$ or $b \to d\,a$ transitions) are not expected, 
being suppressed by two loops and by the relatively large scale $f_a$ typically above $10^6$ GeV (see for example the discussion 
in \cite{Guerrera:2021yss,Guerrera:2022ykl}). 

Majoron couplings with weak gauge bosons arise at one-loop level but with a $\mathcal{O}(1/f_a^2)$ suppression or at two-loops  
at $\mathcal{O}(1/f_a)$, therefore they are not phenomenology appealing. Conversely, $\mathcal{O}(1/f_a)$ two-loop contributions 
to the Majoron-photons couplings can be potentially relevant~\cite{Heeck:2019guh}, contributing to the anomalous Lagrangian term
\begin{equation}
    \sL_{a\gamma\gamma}=-\dfrac{g_{a\gamma\gamma}}{4}aF_{\mu\nu}\widetilde{F}^{\mu\nu}\,.
\end{equation}
A complete calculation of the two-loop Majoron-photons coupling is beyond the scope of this paper. As $R\ll 1$, we can, however, opportunely 
adapt the results of Ref.~\cite{Heeck:2019guh} to reproduce an approximate $g_{a\gamma\gamma}$ suitable in the parameter space 
of interest for our model. Within the same range of validity, the $g_{a\gamma\gamma}$ coupling reads  
\begin{align}
 g_{a\gamma\gamma}\simeq-\dfrac{\alpha_\text{em}}{8\pi^3 f_a}\left[\Tr{K}\sum\limits_f N_c^f Q_f^2 T_3^f\,h 
  \left(\dfrac{m_a^2}{4m_f^2}\right)+\sum\limits_{\ell=e,\mu,\tau}K_{\ell\ell}\,h\left(\dfrac{m_a^2}{4m_\ell^2}\right)\right]\,, 
  \label{eq:gagaga1}   
\end{align}
that is the result of \cite{Heeck:2019guh} with a global minus sign indicating that RH leptons are predominantly coupled with 
$\phi^*$. In Eq.~\eqref{eq:gagaga1} $f$ runs over all fermions, $N_c$ is the number of colours, $T_3^f$ is the weak isospin and
\begin{equation}
    \label{eq:2loop}
    h(x)\equiv -\dfrac{1}{4x}\left[\log\left(1-2x+2\sqrt{x(x-1)}\right)\right]^2-1 \simeq \begin{cases}
        \, \dfrac{x}{3} & \text{for } x \to 0 \\
        \, -1 & \text{for } x \to \infty
    \end{cases} 
\end{equation}
The dependence of the loop function $h(x)$ implies that the largest contributions come from the lightest generation of fermions, 
namely from electrons and up/down quarks. Furthermore, the anomalous coupling to photons (as well as to gluons) vanishes in the 
$m_a\to 0$ limit, showing that the lowest order amplitude originates from the $\Box a F\tilde{F}$ effective operator.

\begin{figure}[t]
\centering
\includegraphics[width=0.7\textwidth]{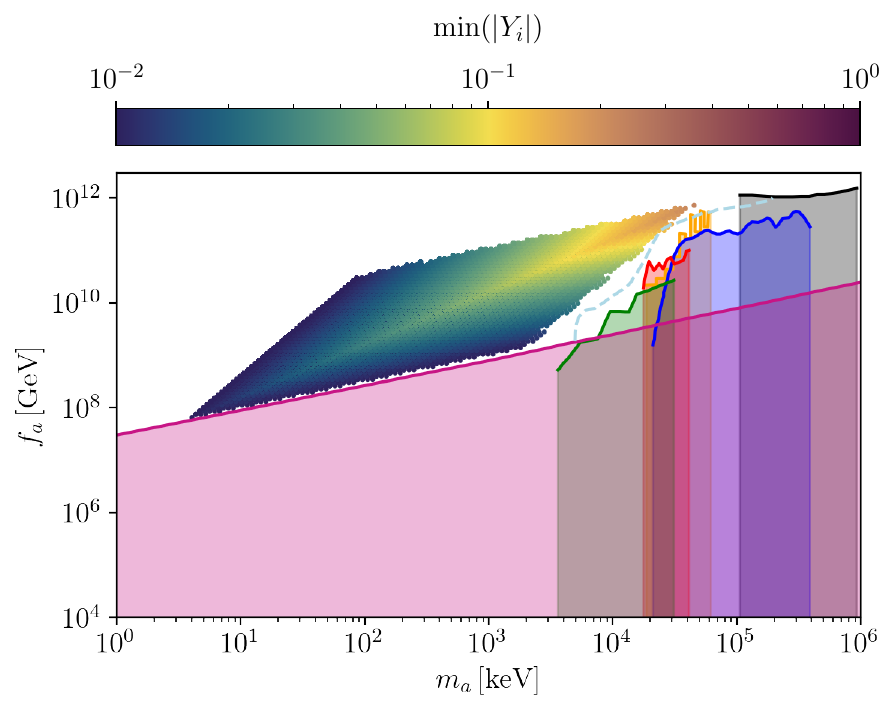}
\caption{\small 
Constraints on Majoron DM. The DM bounds are mainly taken from Ref.~\cite{Akita:2023qiz} and include CMB (\textit{purple area}), 
neutrino experiments, namely Borexino (\textit{green area}), KamLand (\textit{red area}), SK (\textit{blue area}) and projected JUNO (20yr) 
sensitivity (\textit{dashed blue line}). Bounds from Ref.~\cite{Palomares-Ruiz:2007egs}, that include reinterpreted SK data 
(\textit{orange area}) and atmospheric neutrinos data (\textit{gray area}) are also included.}
\label{fig:DM_Majoron}
\end{figure}

%
%
\paragraph{Bounds from Majoron-active neutrino coupling.}
Given the lightness and the feebly interacting nature of the Majoron, it constitutes an appealing candidate for DM~\cite{Akhmedov:1992et, 
Rothstein:1992rh,Berezinsky:1993fm,Lattanzi:2007ux, Gu:2010ys,Esteves:2010sh,Frigerio:2011in,Boulebnane:2017fxw,Biggio:2023gtm}. An exhaustive analysis of the production of the Majoron relic abundance is beyond the scope of this work. For the rest of this paragraph, it will be assumed that Majoron is almost stable, represents the only DM component and predominantly decays into light neutrinos. The strongest available 
constraints, extracted from Refs.~\cite{Akita:2023qiz, Palomares-Ruiz:2007egs}, are shown in Fig.~\ref{fig:DM_Majoron}, where we refer 
for the detailed labelling. Different colours indicate different $\min (|Y_i|)$ values as shown in the upper bar. As pointed out in 
the literature (see e.g.~\cite{Akita:2023qiz}), from DM-Majoron decays into neutrinos one bounds mainly the SSB scale $f_a$ having 
only a mild dependence on the Yukawa couplings. Notice that by simply requiring the ``naturalness conditions'' introduced in 
Sect.~\ref{sec:mmMlag} (i.e. $\epsilon < 0.01 \times \min(|Y_i|)$ and $|Y_i| \in \left[10^{-2},\,  1\right]$) and taming the loop contribution, our prediction lies just 
above the CMB bounds (purple) and on the left of the current neutrino experiments like SK (blue and orange areas), KamLand (red area) 
and Borexino (green area). Hence, scenarios with ``relaxed naturalness conditions'' are being already ruled out by present neutrino 
data. Future neutrino experiments like JUNO (dashed blue line) could, instead, start probing the region of interest for the mmM model.

\paragraph{Bounds on loop-induced Majoron couplings }

\begin{figure}[t!]
\includegraphics[width=\textwidth]{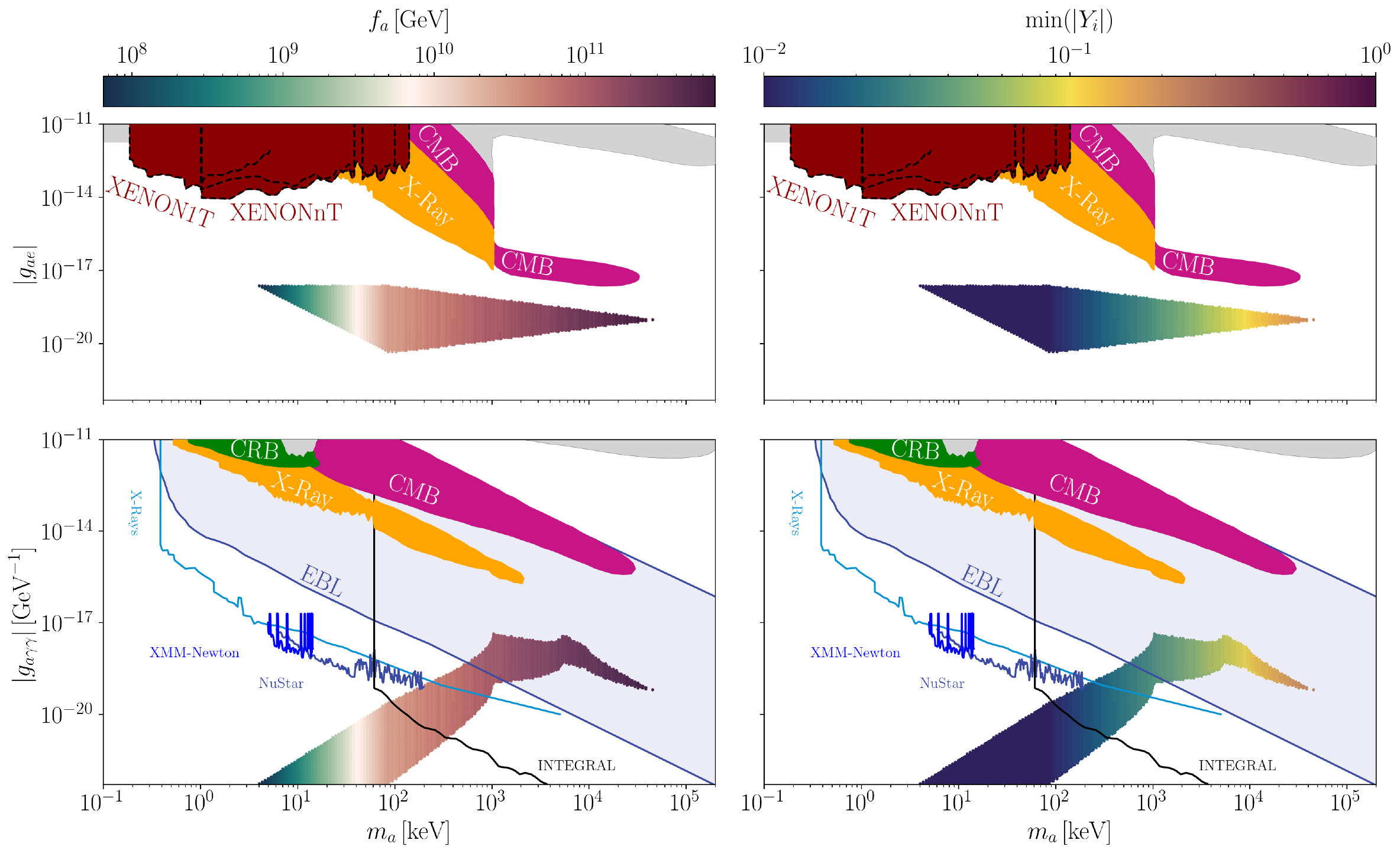}  
\caption{\small  Constraints to Majoron-electron (\textit{top}) and Majoron-photon (\textit{bottom}) couplings. 
Predictions of the mmM model are plotted for different values of the SSB scale $f_a$ (\textit{left}) and 
$\textrm{min}(Y_i)$ (\textit{right}). Irreducible constraints from CMB, CRB and X-Rays (the cyan, green and yellow regions respectively), 
derived from Ref.~\cite{Langhoff:2022bij}, are shown. In the upper plots, the constraints from XENON1T \cite{XENON:2021qze} 
and XENONnT\cite{XENON:2022ltv} (dark red region) are also plotted, while in the lower plots bounds from galactic and extra-galactic 
photons emitted from Majoron decay are depicted, assuming the particle is the full amount of DM.}
\label{fig:ALP_ega}
\end{figure}

Even if the Majoron does not constitute the totality of the observed DM, as assumed in the previous paragraph, it can be bound from other 
astrophysical and cosmological experiments as it would still contribute to the cosmological history of the universe via an irreducible 
freeze-in component \cite{Langhoff:2022bij}. In Fig.~\ref{fig:ALP_ega} (\textit{top}) the bounds on the one-loop Majoron-electron couplings 
of Eq.~\eqref{eq:MajoronLeptonLagrangian} are shown, as a function of the Majoron mass $m_a$ and for different values of the SSB scale 
$f_a$ (\textit{left plot}) and $\min(|Y_i|)$ (\textit{right plot}). Constraints from CMB, CRB and X-Rays (the cyan, green, and yellow regions respectively), 
derived in Ref.~\cite{Langhoff:2022bij}, are shown. As reference the XENON1T \cite{XENON:2021qze} and XENONnT bounds \cite{XENON:2022ltv} 
(darker red) are also plotted. All these data produce constraints that are still at least two orders of magnitude away from the prediction 
of the mmM model represented by the two triangular regions in the lower part of the plots\footnote{We acknowledge the use of the axion bounds repository for these plots \cite{AxionLimits}.}.

Of higher interest are the cosmological bounds that can be derived on the Majoron-photon coupling of Eq.~\eqref{eq:gagaga1}, and that are shown in two lower plots in Fig.~\ref{fig:ALP_ega} as function of the Majoron mass $m_a$ 
and for different values of the SSB scale $f_a$ (\textit{left plot}) and $\min(|Y_i|)$ (\textit{right plot}). The particular shape of the mmM predicted region is due to the functional dependence on the fermion to Majoron mass ratios in the loop function $h(x)$
in Eq.~\eqref{eq:2loop}. In these plots, one can observe that, again, the effects of the irreducible Majoron production in CMB, 
CRB and X-Rays (the red, green and yellow regions respectively), are still far from the regions of interest. 
Stronger bounds to the Majoron-photon coupling can, instead, be obtained from the observation of galactic and Extragalactic Background Light (EBL) derived in Ref.~\cite{Cadamuro:2011fd}, and using XMM-Newton from \cite{Foster:2021ngm}, NuStar \cite{Perez:2016tcq,Ng:2019gch,Roach:2022lgo} and 
INTEGRAL \cite{Calore:2022pks}. These experiments are already able to constrain the upper part of the predicted area at 
the cost, however, of a thermalization temperature of the order of the Planck mass (see the discussion in 
Ref.~\cite{Cadamuro:2011fd}) and only one coupling at a time. 

As can be seen from the plots, $g_{a\gamma\gamma}$ grows with $f_a$. This can be understood from Eq.~\eqref{eq:2loop}, wherein the small $x$ region (below the MeV) $g_{a\gamma\gamma}\propto m_a^2/f_a$ until reaching a ``plateau'' in the large $x$-region. From Eq.~\eqref{eq:MajoronMassExplicit} $m_a$ is proportional to $f_a$, once the neutrino mass is given as an input, hence, $g_{a\gamma\gamma}\propto f_a$. This fictitious dependence, which resembles a non-decoupling effect, stems from the neutrino mass constraint and is not valid for arbitrarily large values of $f_a$: as $m_\nu \sim \epsilon |Y_N| |Y_S| v_\text{EW}^2/f_a$ an increase in $f_a$ must be compensated by an increase in $|Y_{N,S}|$. Once the perturbativity limit $|Y_{N,S}|\leq 1$ is imposed, the artificial growth with $f_a$ cannot continue for arbitrarily large values.

\paragraph{Bounds from LFV}

In the mmM model, as can be seen from Eq.~\eqref{eq:MajoronLeptonLagrangian}, Majoron-charged lepton flavour-violating couplings are 
generated at 1-loop level, opening the possibility to study $\ell_i\to\ell_j a$ processes, which decay width reads
\begin{figure}[tb]
\centering
\includegraphics[width=\textwidth]{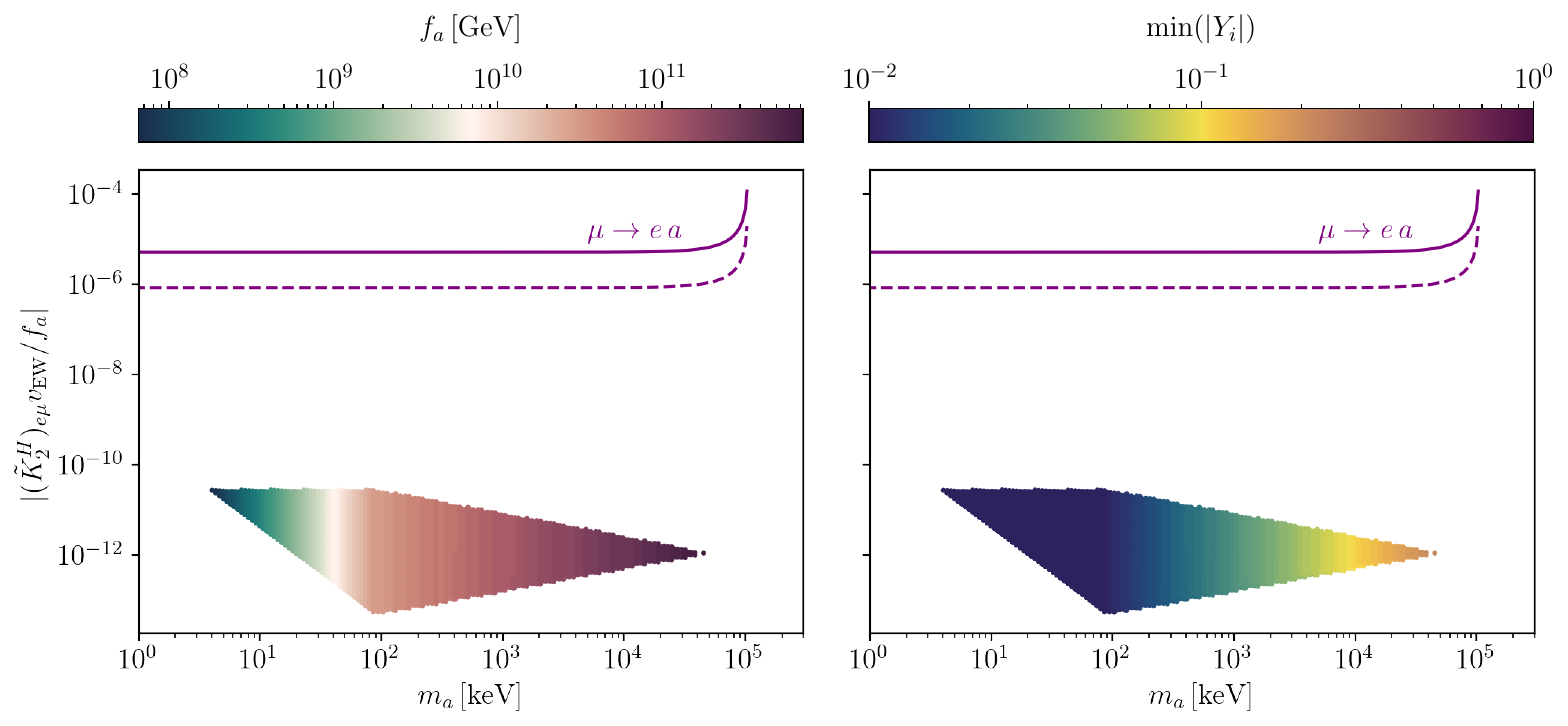}
\caption{\small 
Present (\textit{solid}) and future (\textit{dashed}) 95\%~C.L. bounds on the flavour violating Majoron-e-$\mu$ coupling $(\widetilde{K}^H_2)_{e \mu}$. The mmM model preferred region as functions of the different SSB scale $f_a$ (\textit{left}) and $\min(|Y_i|)$ (\textit{right}) are depicted. }
\label{fig:LFV_Majoron}
\end{figure}
\begin{equation}
\label{eq:LFV-width}
    \Gamma(\ell_i\to\ell_j a)= \dfrac{|(\widetilde{K}^H_2)_{ji}|^2}{2048\pi^5 f_a^2 m_i^3}\sqrt{\lambda(m_i^2,m_a^2,m_j^2)}\left[(m_i^2-m_j^2)^2-m_a^2(m_i^2+m_j^2)\right]\,,
\end{equation}
where $\lambda(x^2,y^2,z^2)$ is the Källén function.

In the Majoron mass range predicted by the mmM model, i.e. $m_a \lesssim 5\times10^4~\text{keV} < m_{\mu}$, the strongest bounds 
come from the $\mu\to e a$ decay, while $\tau\to\mu a$ and $\tau\to e a$ decays would be relevant only for $m_a \gtrsim m_\mu$.
In Fig.~\ref{fig:LFV_Majoron} the constraints from Refs.~\cite{Calibbi:2020jvd,DiLuzio:2023ndz} on the LFV coupling $(\widetilde{K}_2^H)_{e \mu}$ 
defined in Eqs.~\eqref{eq:Ktilde} are shown. We notice that both the present and future sensitivity of LFV experiments are still orders 
of magnitude far from testing the region of interest for the mmM model.

%
%
\section{Conclusions}
\label{sec:Concls}
%
%

The Majoron is traditionally considered as the would-be-Goldstone boson of the spontaneous breaking of the Lepton Number. The origin of its mass is an old problem and in this paper, we propose a scenario where it naturally arises without invoking Planck effects or extra ingredients behind those strictly necessary to correctly describe lepton masses and mixing. 

The model is a high-scale Seesaw mechanism with two right-handed neutrinos with a mass texture that is very similar to the one associated with the so-called linear Seesaw. The heavy neutral lepton masses arise after the spontaneous symmetry breaking of a global Abelian symmetry, while the active neutrino masses can be correctly described only by introducing an explicit breaking of such symmetry.
The Majoron is generated after the spontaneous breaking and acquires mass due to the explicit breaking. Although this may appear to be completely generic, we showed that it is not: not all the possible explicit breakings of the symmetry would lead to a mass for the Majoron. 

We identified the unique, minimal model where the Majoron mass and the active neutrino masses are strictly tied together, by requiring three conditions to be satisfied: renormalisability of the model Lagrangian; only one Majorana scale that is associated with the spontaneous breaking of the Abelian symmetry; only one explicit breaking of this symmetry. The latter thus plays the role of both Lepton Number and PQ symmetry. 

As a consequence of these conditions, only one scalar field $\phi$ transforming under the Abelian symmetry is introduced in the spectrum and, to guarantee a mass for the Majoron, both $\phi$ and $\phi^\ast$ enter the Lagrangian. In particular, this scalar field appears in the Majorana terms of the Lagrangian that represent the mass terms for the heavy neutral leptons, once it develops a vacuum expectation value. If these Majorana terms constitute the spontaneous symmetry-breaking sector, one of the Dirac terms explicitly violates the symmetry. The resulting neutral lepton mass matrix, after both LN/PQ and electroweak symmetry breaking, has already been studied in the pure neutrino context and undergoes the name of Extended Seesaw limit: it predicts potentially large one-loop contributions to the active neutrino masses that are kept under control only focusing in a parameter space where the two heavy neutral leptons have almost degenerate masses. 

All in all, active neutrino masses and the lepton mixing can be correctly described with heavy neutral leptons with masses in the range $[10^7,\,10^{12}]\GeV$, when the Dirac and Majorana Yukawa couplings span the range $[10^{-2},\,1]$, for both the normal and inverse ordering of the active neutrino mass spectrum. 

The Majoron mass arises at one-loop and proportional to the active neutrino masses. We performed the explicit computation and confirmed the result through the CW potential, in the $\ov{\text{MS}}$-scheme. We found that the derivative or chirally preserving basis for the Majoron couplings greatly simplifies the computations. In our model, the Majoron mass turns out to scale logarithmically with the spontaneous symmetry breaking scale $f_a$. This is in contrast with the traditional QCD axion models where the mass is inversely proportional to $f_a$. Once inserting neutrino data as inputs and requiring natural values for the Yukawa couplings $\in[10^{-2},\,1]$, we obtain a Majoron mass in the range $[1,\,10^{5}]\keV$ corresponding to $f_a\in[10^8,\,10^{12}]\GeV$.

This light and feebly interacting Majoron represents an appealing candidate for Dark Matter. We performed phenomenological analyses investigating the impact of its couplings with active neutrinos, charged leptons and photons, the last two being loop-induced. Majoron-neutrino interactions have an impact in the CMB and in oscillation neutrino experiments: the proper parameter space of the model lays just next to the excluded region by these experiments and therefore it would be directly probed once the experimental uncertainties decrease. Majoron couplings with electrons and photons are tested by CMB, CRB and X-Rays, but their sensitivities do not reach the proper parameter space of our model. The same holds for XENON1T and XENONnT (expected) bounds. For the specific case of the couplings with photons, the model could be probed from the observation of galactic and extra-galactic photon spectrum. Interestingly, the Majoron-photon coupling grows linearly with $f_a$ in the examined parameter space, and therefore the first region that could be tested corresponds to the highest values of $f_a$ and largest Majoron masses. Finally, our Majoron describes lepton flavour-violating processes, $\mu\to e a$, $\tau\to\mu a$ and $\tau\to ea$, but far from testing the proper parameter space of the model.  

To summarise, the minimal massive Majoron Seesaw model represents the first example where the Majoron mass naturally arises from the same context where active neutrino masses and the lepton mixing are correctly described, without the necessity of introducing {\it ad hoc} new parameters or invoking high-scale-suppressed operators e.g.~Planck-induced contributions. Its minimality conditions imply the uniqueness of the model and link the Majoron mass to the active neutrino masses providing a phenomenological viable scenario that could be probed by CMB, EBL, and oscillation neutrino experiments in the near future. 


%
\section*{Acknowledgements}

The authors thank Ilaria Brivio, Luca Di Luzio, Bel\'en Gavela, Manuel Gonz\'alez L\'opez, Álvaro Lozano Onrubia, 
Jorge Dasilva Golan, 
Jonathan Machado Rodr\'iguez, Daniel Naredo Tuero and Giuseppe Lucente for useful discussions. The authors also thank Avelino Vicente and Antonio Herrero-Brocala for spotting an error in an equation of the previous version. AdG thanks Carlos 
A. Argüelles and the group of Palfrey House for their hospitality and the stimulating working environment during which a core part 
of this work was realised. AdG also thanks J. Jaeckel and the Institute for Theoretical Physics of the University Heidelberg for relevant remarks to this work and the warm hospitality. The authors acknowledge partial financial support by the European Union's Horizon 2020 research and 
innovation programme under the Marie Sk\l odowska-Curie grant agreements No 860881-HIDDeN and 101086085-ASYMMETRY. AdG, LM and SR  
acknowledge partial financial support by the Spanish Research Agency (Agencia Estatal de Investigaci\'on) through the grant 
IFT Centro de Excelencia Severo Ochoa No CEX2020-001007-S and by the grants PID2019-108892RB-I00 and PID2022-137127NB-I00 funded by MCIN/AEI/ 10.13039/501100011033. 
The work of AdG and XPD was supported by the European Union's Horizon 2020  Marie Sk\l odowska-Curie grant agreement No 860881-HIDDeN. This article/publication is based upon work from COST Action COSMIC WISPers CA21106, supported by COST (European Cooperation in Science and Technology). This work was also partially supported by the Italian MUR Departments of 
Excellence grant 2023-2027 ``Quantum Frontiers''. 

\appendix
%
%
\section{ALP PQ-Breaking Interactions}
\label{app:MassEigen}
%
%
In this appendix, we give some details regarding the Majoron PQ-breaking interactions in the lepton mass basis. This is necessary to calculate the Majoron mass in Section \ref{sec:MajoronMass}. After rotating away the PQ-preserving terms, the only term relevant for the Majoron mass is
\be
-\sL_{a}\supset\epsilon\, \ov{L_L}\,\tH\,Y_S\,S_R\,e^{2ia/f_a}+\hc\,.
\label{RelevantTerm}
\ee
The block diagonalisation of the neutral mass matrix $\cM_\chi$ can be performed through a unitary redefinition of the field vector $\chi_L$~\cite{Blennow:2011vn}. As we are interested in interactions of $\mathcal{O}(\hat{\Lambda}^{-1})$, we expand the unitary matrix at second order in the mixing matrix $\Theta$
\begin{equation}
	\chi_L\rightarrow U_\chi\,\chi_L\qquad\text{with}\qquad U_\chi\simeq 
	\begin{pmatrix}
		\unity-\dfrac{1}{2}\Theta\Theta^\dagger &\Theta\\
		-\Theta^\dagger &\unity-\dfrac{1}{2}\Theta^\dagger\Theta
	\end{pmatrix}\,,
\end{equation}
which requires at this order
\begin{equation}
	\begin{split}
		\Theta\simeq \hat{m}\, \hat{\Lambda}^{-1}&=
		\begin{pmatrix}
			m_N & \epsilon\,m_S
		\end{pmatrix}\begin{pmatrix}
		\Lambda_{NN} & \Lambda_{NS}\\
		\Lambda_{NS} & 0
	\end{pmatrix}^{-1}=
	\dfrac{1}{\Lambda_{NS}}\begin{pmatrix}
		\epsilon\,m_S & m_N-\epsilon\,m_S\dfrac{\Lambda_{NN}}{\Lambda_{NS}}
	\end{pmatrix}\,.
\end{split}
\end{equation}
After this first diagonalisation, the fields get rotated, at LO in $\Theta$, as
\begin{equation}
\label{eq:field-redefinitions}
	\begin{cases}
		\hspace{0,5cm}\nu_L &\rightarrow \quad
		\nu_L+\Theta \begin{pmatrix}
			N_R^c\\
			S_R^c
		\end{pmatrix}=\nu_L+ \dfrac{\epsilon\,m_S}{\Lambda_{NS}}  N_R^c + \dfrac{1}{\Lambda_{NS}}\left(m_N-\epsilon\, m_S \dfrac{\Lambda_{NN}}{\Lambda_{NS}}\right)S_R^c\,,\\
		\begin{pmatrix}
			N_R^c\\
			S_R^c
		\end{pmatrix} &\rightarrow\quad
		\begin{pmatrix}
			N_R^c\\
			S_R^c
		\end{pmatrix}-\Theta^\dagger\,\nu_L=
		\begin{pmatrix}
			N_R^c\\
			S_R^c
		\end{pmatrix}-
		\dfrac{1}{\Lambda_{NS}}\begin{pmatrix}
			\epsilon\,m_S^\dagger\\
			m_N^\dagger-\epsilon\,m_S^\dagger\dfrac{\Lambda_{NN}}{\Lambda_{NS}} 
		\end{pmatrix}\nu_L\,.
	\end{cases}
\end{equation}
On the other side, the HNLs are not yet written in terms of mass eigenstates: we still need to diagonalise the $23$ sector of the resulting mass matrix after the block diagonalisation. To do so, we perform a second field redefinition that only affects the neutral exotic states:
	\begin{equation}
 \label{eq:rotation-heavy}
		\begin{pmatrix}
			N_R & S_R
		\end{pmatrix}^T \rightarrow U_N\begin{pmatrix}
		N_R & S_R
	\end{pmatrix}^T
	\qquad\text{with}\qquad
	U_N=\dfrac{1}{\sqrt{\,M_N+M_S}}\begin{pmatrix}
		i\sqrt{M_N}&\sqrt{M_S}\\[1mm]
		-i\sqrt{M_S}&\sqrt{M_N}
	\end{pmatrix}\,,
\end{equation}
obtaining the HNLs masses in Eq.~\eqref{eq:HNL-masses}.
The parameters $\Lambda_{NS,NN}$ can be written in terms of the HNL masses via
\begin{align}
	&\Lambda_{NS}=\sqrt{M_N M_S}\,, & \Lambda_{NN}=M_S-M_N\,.
\end{align}
Combining these field redefinitions we obtain the explicitly breaking term in the lepton mass basis:
\begin{align}
	-\sL_{a}&\supset  \left(e^{2ia/f_a}-1\right)\left\{-\dfrac{1}{\sqrt{M_N M_S}}\ov{\nu_L}U^\dagger_\text{PMNS}\left(\epsilon m_S m_N^T-\left(\dfrac{\alpha^2-1}{\alpha}\right)\epsilon^2 m_S m_S^T\right)U^\ast_\text{PMNS}\nu_L^c\right.\\
	&\nn-\dfrac{i}{\sqrt{(M_N+M_S)}}\left(\sqrt{M_S}\ov{\nu_L}U^\dagger_\text{PMNS}\epsilon m_S N_R-\sqrt{M_N}\ov{\nu_L}U^\dagger_\text{PMNS}\epsilon m_S S_R\right)\\
	&+\nn(\ov{N_R^c}N_R)\dfrac{1}{(M_N+M_S)}\left[\epsilon^2 m_S^\dagger m_S-\alpha\left(\epsilon m_N^\dagger m_S -\left(\dfrac{\alpha^2-1}{\alpha}\right)\epsilon^2 m_S^\dagger m_S\right)\right]\\
	&+\nn(\ov{S_R^c}S_R)\dfrac{1}{(M_N+M_S)}\left[\epsilon^2 m_S^\dagger m_S+\dfrac{1}{\alpha}\left(\epsilon m_N^\dagger m_S -\left(\dfrac{\alpha^2-1}{\alpha}\right)\epsilon^2 m_S^\dagger m_S\right)\right]\\
	&+\nn(\ov{N_R^c}S_R) \dfrac{i}{(M_N+M_S)}\left[\dfrac{1}{\alpha}\epsilon^2 m_S^\dagger m_S-\left(\epsilon m_N^\dagger m_S -\left(\dfrac{\alpha^2-1}{\alpha}\right)\epsilon^2 m_S^\dagger m_S\right)\right]\\
	&\left.-\nn(\ov{S_R^c}N_R) \dfrac{i}{(M_N+M_S)}\left[\alpha\epsilon^2 m_S^\dagger m_S+\left(\epsilon m_N^\dagger m_S -\left(\dfrac{\alpha^2-1}{\alpha}\right)\epsilon^2 m_S^\dagger m_S\right)\right]\right\}\\
	&\nn+\hc\,,
	\label{ChiPreservingLagIntStep}
\end{align}
where we defined $\alpha\equiv \sqrt{M_S/M_N}$. The derivative couplings can be obtained similarly by employing the field redefinitions of Eq.~\eqref{eq:field-redefinitions} in the Lagrangian of Eq.~\eqref{MajoronLagChiPreserving} and are not going to be explicitly reported here.

Not all the interactions presented above are relevant to the LO prediction of the Majoron mass. First of all, as we are in the chirality preserving basis, 
only the Balloon diagram can provide relevant contributions at LO. This prompts us to neglect any term with powers of $\epsilon^2$ and consider only diagonal couplings, leaving us with
	\be
	\label{ChiPreservingBasisComplete} 
	-\sL_{a}\supset\left[\dfrac{1}{2}\ov{\nu_L}m_\nu\nu_L^c+\dfrac{2\epsilon\,m_N^\dagger m_S}{\sqrt{M_N\,M_S}
		(M_N+M_S)}\Bigg(M_N\ov{S_R^c}S_R-M_S\ov{N_R^c}N_R\Bigg)\right]e^{2ia/f_a}+\hc\,,
	\ee
	Moreover, as concluded in Eq.~\eqref{toymodelResultsChiFlipping} for 
	the toy model, such contributions are proportional to the fermion running in the loop and therefore we can focus on the HNL couplings 
	in Eq.~\eqref{ChiPreservingBasisComplete}. The mixed $\ov{S_R^c}N_R$ term only contributes to the sub-leading Bubble diagram, while the active neutrino term contribution to the dominant Balloon diagram is proportional to the light neutrino masses, being therefore completely negligible. 
	We can thus restrict our considerations 
	to the following terms of the Lagrangian
	\begin{equation}
		\begin{split}
			-\sL_{a}\supset
   &\dfrac{2\epsilon\,m_N^\dagger m_S}{\sqrt{M_N\,M_S}(M_N+M_S)}\Bigg(M_N\ov{S_R^c}S_R-M_S\ov{N_R^c}N_R\Bigg)\dfrac{a^2}{f_a^2}+\hc\,.
		\end{split}
		\label{FinalChiPreservingBasisRelevant}
	\end{equation}

%
%

\section{Majoron and HNLs Interactions}
\label{app:Majoron-interactions}
The interaction matrix with the Majoron can be compactly written down in block matrix form. After the block diagonalization has been carried out, to move to the mass basis, one must include the unitary matrix
\begin{equation}
    U\equiv \begin{pmatrix}
        U_\text{PMNS} & 0
        \\
        0 & U_N
    \end{pmatrix}\,,
\end{equation}
where $U_\text{PMNS}$ is the light neutrinos PMNS matrix and $U_N$ is defined in Eq.~\eqref{eq:rotation-heavy}. The interactions in the mass basis reads
\begin{equation}
    \sL_a\simeq -\dfrac{1}{2}\ov{\chi_L}U^\dagger \mathcal{M}_a U^\ast\chi_L^c\,.
\end{equation}
where
\begin{equation}
    \mathcal{M}_a=\begin{pmatrix}
        (\mathcal{M}_a)_{11} & (\mathcal{M}_a)_{12}\\
        (\mathcal{M}_a)_{21} &  (\mathcal{M}_a)_{22}
    \end{pmatrix}
\end{equation}
and its components are given by
\begin{align}
    &(\mathcal{M}_a)_{11} \simeq -m^{\text{TL}}_\nu e^{i\sigma a/fa}+\epsilon^2 m_S m_S^T \dfrac{\Lambda_{NN}}{\Lambda_{NS}}\left(e^{ia/f_a}-e^{i\sigma a/f_a}\right)\,,\\
    &(\mathcal{M}_a)_{12}=[(\mathcal{M}_a)_{21}]^T \simeq \begin{pmatrix}
        -e^{i\sigma a/f_a}m_N+\left(e^{ia/f_a}-e^{i\sigma a/f_a}\right)\epsilon m_S\dfrac{\Lambda_{NN}}{\Lambda_{NS}} & \epsilon m_S e^{i\sigma a/f_a}
    \end{pmatrix}\,,\\
    &(\mathcal{M}_a)_{22}\simeq \begin{pmatrix}
        \Lambda_{NN}e^{ia/f_a} & \Lambda_{NS}e^{i\sigma a/f_a}\\
        \Lambda_{NS} e^{i\sigma a/f_a} & 0
    \end{pmatrix}\,,
\end{align}
where we have introduced a parameter $\sigma$ that helps us interpolate between the typical Majoron of the literature, $\sigma=+1$ all mass terms couple to $\phi$, and other non-standard scenarios such as the mmM, where $\sigma=-1$, $\Lambda_{NS}$ couples to $\phi^\ast$ while $\Lambda_{NN}$ to $\phi$. 
Expanding the exponential at LO, we thus have
\begin{align}
\label{eq:Lagmassbasis}
    &(\mathcal{M}_a)_{11} \simeq \dfrac{ia}{f_a}\left[ -\sigma m^{\text{TL}} +\epsilon^2 m_S m_S^T \dfrac{\Lambda_{NN}}{\Lambda_{NS}}\left(1-\sigma\right)\right]\,,\\
    &(\mathcal{M}_a)_{12}=[(\mathcal{M}_a)_{21}]^T \simeq  \dfrac{ia}{f_a}\begin{pmatrix}
       -\sigma m_N+\left(1-\sigma\right)\epsilon m_S\dfrac{\Lambda_{NN}}{\Lambda_{NS}} & \sigma \epsilon m_S 
    \end{pmatrix}\,,\\
    &(\mathcal{M}_a)_{22}\simeq \dfrac{ia}{f_a}\begin{pmatrix}
        \Lambda_{NN} & \sigma\Lambda_{NS}\\
        \sigma\Lambda_{NS}  & 0
    \end{pmatrix}\,.
\end{align}

The interactions with gauge bosons can be derived similarly and in the mass basis they read
\begin{align}
    &\sL_{Z}\supset -\dfrac{g}{2\cos\theta_W}\ov{\chi_L}\slashed{Z}\,U^\dagger \begin{pmatrix}
        1-\Theta\Theta^\dagger & \Theta\\
        \Theta^\dagger & \Theta^\dagger\Theta
    \end{pmatrix}U\,\chi_L\,,\\
    &\sL_{W^\pm}\supset -\dfrac{g}{\sqrt{2}} \ov{\chi_L}\slashed{W}^+\,U^\dagger\begin{pmatrix}
        1-\dfrac{1}{2}\Theta\Theta^\dagger & 0\\
        \Theta^\dagger & 0
    \end{pmatrix}\ell_L\,,\\
    &\sL_h\supset -\left(\dfrac{h}{v}\right)\ov{\chi_L}\,U^\dagger \begin{pmatrix}
        \hat{m}_\nu &  \dfrac{1}{2}\hat{m}\\
        \dfrac{1}{2}\hat{m}^T & \dfrac{1}{2}\left(\Theta^\dagger \hat{m}+ \hat{m}^T\Theta^\ast\right)
    \end{pmatrix}U^\ast\,\chi_L^c\,,
\end{align}
where recall that 
\begin{equation}
	\begin{split}
		\Theta\simeq \hat{m}\, \hat{\Lambda}^{-1}
		=
	\dfrac{1}{\Lambda_{NS}}\begin{pmatrix}
		\epsilon\,m_S & m_N-\epsilon\,m_S\dfrac{\Lambda_{NN}}{\Lambda_{NS}}
	\end{pmatrix}\,,
\end{split}
\end{equation}
and thus
\begin{align}
    &\Theta\Theta^\dagger \simeq\dfrac{1}{\Lambda_{NS}^2}\left[\epsilon^2 m_S m_S^\dagger +\left(m_N -\epsilon m_S\dfrac{\Lambda_{NN}}{\Lambda_{NS}}\right)\left(m_N^\dagger -\epsilon m_S^\dagger\dfrac{\Lambda_{NN}}{\Lambda_{NS}}\right)\right]\,,\\
    &\Theta^\dagger\Theta \simeq\dfrac{1}{\Lambda_{NS}^2}\begin{pmatrix}
       \epsilon^2 m_S^\dagger m_S & \epsilon m_S^\dagger \left(m_N -\epsilon m_S\dfrac{\Lambda_{NN}}{\Lambda_{NS}}\right)\\
       \left(m_N^\dagger -\epsilon m_S^\dagger\dfrac{\Lambda_{NN}}{\Lambda_{NS}}\right) \epsilon m_S & \left(m_N^\dagger -\epsilon m_S^\dagger\dfrac{\Lambda_{NN}}{\Lambda_{NS}}\right)\left(m_N -\epsilon m_S\dfrac{\Lambda_{NN}}{\Lambda_{NS}}\right)
    \end{pmatrix}\,.
\end{align}

%
%
\section{One--Loop Effective Potential and Majoron Mass}
\label{app:CW}
%
%
In this appendix we provide an alternative derivation of the Majoron mass, computing the full one--loop contribution to the scalar potential through the Coleman-Weinberg (CW) potential~\cite{Coleman:1973jx}. Accounting already for the trace over the Dirac indices, the fermionic contribution, in 
the $\overline{\mbox{MS}}$ scheme, reads
\be
V_\text{CW}=
-\dfrac{1}{2}\times\dfrac{1}{16\pi^2}\Tr\left[\left(\cM_\chi\cM_\chi^\dagger\right)^2\left(
\log\left(\dfrac{\cM_\chi\cM_\chi^\dagger}{\mu_R^2}\right)-\frac{3}{2}\right)\right]
\label{eq:CWMSb}
\ee
where $\cM_\chi\equiv\cM_\chi(H,\phi)$ is the neutral mass matrix in Eq.~\eqref{MajoronNeutralMassMatrix} including the dependence of the 
scalar fields $H$ and $\phi$. In the following a compact notation for the neutral mass matrix $M_\chi$ is going to be adopted:
\be
\mathcal{M}_\chi(H,\phi)=\begin{pmatrix}
	0 & \widehat{m}(H)\\
	\widehat{m}(H)^T & \widehat{\Lambda}(\phi)
\end{pmatrix}\,,
\label{CVBlockNotation}
\ee
where $\widehat{m}$ and $\widehat{\Lambda}$ are the field-dependent Dirac and Majorana blocks inheriting the structure of the mass matrix in 
Eq.~\eqref{MajoronNeutralMassMatrix}. The overall factor $1/2$ in front of the expression is due to the Majorana nature of the fermionic fields involved, which have half of the degrees of freedom of Dirac fermions. 

We turn now to the explicit computation of the relevant traces. We consider only the terms containing the Majoron contribution. We find
\begin{align}
	\Tr\left[(\mathcal{M}_\chi \mathcal{M}_\chi^\dagger )^2\right]&=2\Re\Big(\Tr\left[(\widehat{m}^\dagger\widehat{m})^2\Big]\right)+
	\Tr\left[(\widehat{\Lambda}^\dagger\widehat{\Lambda})^2\right]+2\left(\Tr\left[\widehat{m}^\dagger\widehat{m}\widehat{\Lambda}^\dagger
	\widehat{\Lambda}\right]+\Tr\left[\widehat{m}^T\widehat{m}^\ast\widehat{\Lambda}\widehat{\Lambda}^\dagger\right]\right)\nn\\
	&\supset2\,\epsilon\,Y_{NS}\,Y_{NN}\,v_\text{EW}^2\left(Y_N^\dag\,Y_S\,\phi^2+Y_S^\dag\,Y_N\,\phi^{\ast2}\right)
	\label{TrM4}\\	&\supset2\,\epsilon\,Y_{NS}\,Y_{NN}\,|Y_N|\,|Y_S|\,|\eta|\,v_\text{EW}^2\,f_a^2\,\cos\left(\vartheta_\eta+\dfrac{2a}{f_a}\right)\,,\nn
\end{align}
where in the last step we made use of the definitions in Eq.~\eqref{etaDEF}. This confirms that the Majoron mass can only obtain a contribution 
when the four $Y_i$ couplings are present and non-vanishing, and the leading contribution turns out to be linear in the LN breaking parameter $\epsilon$. 
The computation of the log is far more problematic as one needs the eigenvalues of the full $5\times 5$ matrix, $\mathcal{M}_\chi \mathcal{M}_\chi^\dagger$. However, recalling from the explicit computation that the one-loop diagrams contributing to the Majoron mass are proportional to internal lepton masses, we can safely neglect in the calculation the light active neutrino masses. Denoting with $\{\mu_i\}_{i=1}^5$ the eigenvalues of $\mathcal{M}_\chi \mathcal{M}_\chi^\dagger$, this allows to write
\begin{align}
	&\Tr{\mathcal{M}_\chi \mathcal{M}_\chi^\dagger}\approx \mu_1+\mu_2\,, & \Tr{(\mathcal{M}_\chi \mathcal{M}_\chi^\dagger)^2}\approx \mu_1^2+\mu_2^2\,,
\label{eq:eigensys}
\end{align}
where $\mu_{1,2}$ represent the two large HNL masses. By solving the system in Eq.~\eqref{eq:eigensys} the CW potential for the Majoron field, in the $\overline{\mbox{MS}}$ scheme reads:
\be
\begin{split}
	V_\text{CW}=&-\epsilon\dfrac{Y_{NS}\,Y_{NN}\,|Y_N|\,|Y_S|\,|\eta|\,v_\text{EW}^2\,f_a^2}{16\pi^2}\,
	\cos\left(\vartheta_\eta+\dfrac{2a}{f_a}\right)\times\\
	& \times\Bigg[\dfrac{(Y_{NN}^2+2Y_{NS}^2)}{Y_{NN} \sqrt{Y_{NN}^2+4Y_{NS}^2}}\text{Arcoth}\left(\dfrac{(Y_{NN}^2+2Y_{NS}^2)}{Y_{NN} 
		\sqrt{Y_{NN}^2+4Y_{NS}^2}}\right)+\left(\log\left(\dfrac{Y_{NS}^2 f_a^2}{2\mu^2_R}\right)-1\right)\Bigg]\,.
	\label{TrM4logM2}
\end{split}
\ee
To obtain the final expression one has to use of the well-known algebraic identity:
\be
\text{Arcoth}(x)=\dfrac{1}{2}\log{\frac{x+1}{x-1}}\,.
\ee

A few comments are in order. First of all, expanding the cosine, we would get a linear term for the Majoron, corresponding to a tadpole. To avoid it, we can perform a shift in the Majoron field
\be
\dfrac{2a}{f_a}\rightarrow\dfrac{2a}{f_a}-\vartheta_\eta\,,
\label{MajoronRedefThetaeta}
\ee
with a consequent appearance of $e^{\pm i\vartheta_\eta/2}$ term in Eq.~\eqref{eq:neutrinosectortot}. As already mentioned, the couplings 
$Y_{NS}$ and $Y_{NN}$ can be made real by a proper redefinition of the leptons fields and, as a result, the $\vartheta_\eta$ phase would 
eventually end up in the Dirac Yukawa couplings. This is more evident in the chirality conserving basis in Eq.~\eqref{MajoronLagChiPreserving}, 
where the only term that would be affected by the Majoron shift is the LN explicit breaking term: the $\vartheta_\eta$ phase can be absorbed 
in the definition of $Y_S$,
\be
Y_S\,e^{-i\vartheta_\eta}\rightarrow Y_S\,.
\ee
The net effect of this redefinition propagates to Eq.~\eqref{etaDEF}, where the parameter $\eta$ turns out to be a real number. This resembles 
what occurs with the QCD axion, $a_\text{QCD}$. As discussed by Vafa and Witten~\cite{Vafa:1983tf}, the QCD vacuum energy has its minimum 
when $a_\text{QCD}\rightarrow a_\text{QCD}-\ov\theta$, where $\ov\theta$ includes a phase from the quark mass matrices. The latter, which 
is physical only in the presence of the axion, gets fixed by a minimum condition that also prevents the appearance of the axion tadpole. 
As the HNL sector is not coupled to SM gauge bosons, this shift does not induce any anomalous SM gauge boson coupling. 

By taking the second derivative on the CW potential, one automatically obtains the one-loop contribution to the Majoron mass:
\be
\begin{split}
	m^2_a=&\,\epsilon\,\dfrac{Y_{NS}\,Y_{NN}\,|Y_N|\,|Y_S|\,|\eta|\,v_\text{EW}^2}{4\pi^2}\times\\
	&\times\Bigg[\dfrac{(Y_{NN}^2+2Y_{NS}^2)}{Y_{NN} \sqrt{Y_{NN}^2+4Y_{NS}^2}}\text{Arcoth}\left(\dfrac{(Y_{NN}^2+2Y_{NS}^2)}{Y_{NN} 
		\sqrt{Y_{NN}^2+4Y_{NS}^2}}\right)+\left(\log\left(\dfrac{Y_{NS}^2 f_a^2}{2\mu^2_R}\right)-1 \right)\Bigg]\,.
  \label{MajoronMassCW}
\end{split}
\ee

%
\footnotesize
\bibliography{bibliography}{}
\bibliographystyle{BiblioStyle}

\end{document}